# Directed Growth of Hydrogen Lines on Graphene: High Throughput Simulations Powered by Evolutionary Algorithm

G. Özbal,[1] J. T. Falkenberg,[2,3] M. Brandbyge,[2,3] R. T. Senger,[1] and H. Sevinçli[4,*]

[1]*Physics Department, Izmir Institute of Technology, 35430 Izmir, Turkey.*
[2]*Department of Micro- and Nano-technology, Technical University of Denmark, DK-2800 Kongens Lyngby, Denmark*
[3]*Center for Nanostructured Graphene(CNG), Department of Micro- and Nano-technology, Technical University of Denmark, DK-2800 Kongens Lyngby, Denmark*
[4]*Department of Materials Science and Engineering, Izmir Institute of Technology, 35430 Izmir, Turkey.*

We set up an evolutionary algorithm combined with density functional tight-binding (DFTB) calculations to investigate hydrogen adsorption on flat graphene and graphene monolayers curved over substrate steps. During the evolution, candidates for the new generations are created by adsorption of an additional hydrogen atom to the stable configurations of the previous generation, where a mutation mechanism is also incorporated. Afterwards a two-stage selection procedure is employed. Selected candidates act as the parents of the next generation. The evolutionary algorithm predicts formation of lines of hydrogen atoms on flat graphene. In curved graphene, the evolution follows a similar path except for a new mechanism, which aligns hydrogen atoms on the line of minimum curvature. The mechanism is due to the increased chemical reactivity of graphene along the minimum radius of curvature line (MRCL) and to $sp^3$ bond angles being commensurate with the kinked geometry of hydrogenated graphene at the substrate edge. As a result, the reaction barrier is reduced considerably along the MRCL, and hydrogenation continues like a mechanical chain reaction. This growth mechanism enables lines of hydrogen atoms along the MRCL, which has the potential to overcome substrate or rippling effects and could make it possible to define edges or nanoribbons without actually cutting the material.

One of the most attractive features of two-dimensional materials is their ability to be functionalized in more effective ways than conventional materials. However graphene lacks an energy band gap, a requirement for graphene to be useful in digital electronics applications. Adsorption of hydrogen atoms is a direct way to alter chemical and physical properties of graphene[1] and hydrogenation in a controlled way was shown to open a band gap.[2–9]. There are proposals to tune the band gap of graphene by hydrogen adsorbed nanoripples[10–12] and forming superlattices consisting of lines of adsorbed hydrogen atoms.[13,14] It is well known that curvature enhances chemical reactivity of graphene and related materials by decreasing the activation barrier for adsorbates.[15–22] This property was used to enhance hydrogen storage capacity of graphene and CNTs.[23–27] A straightforward way to deform graphene is to introduce steps in the underlying substrate. Such deformation also affects electronic and thermal transport properties.[28–30]

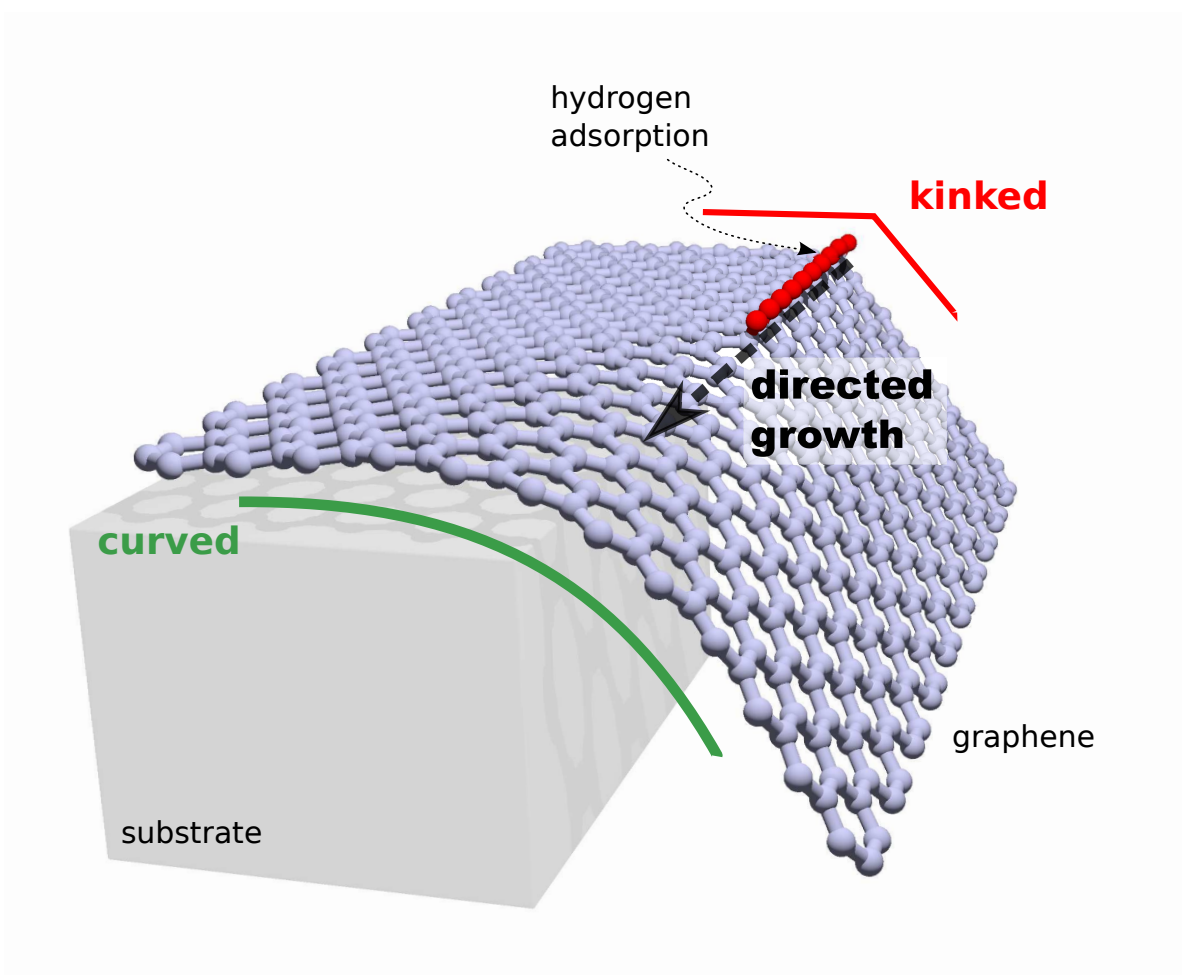

FIG. 1. Illustration of the directed growth of hydrogen atoms on the minimum radius of curvature line. Propagation of the hydrogen atoms along the first and the second minimum radius of curvature lines leads to formation of a kink line.

Stability of small hydrogen clusters on single and multilayer graphene, as well as on carbon nanotubes have been studied by several groups theoretically and experimentally.[17,31–40] Also there have been attempts to determine the most stable graphane-like clusters and to control the size of the hydrogen islands.[41,42] Those studies were focused either on individual dimer and trimer configurations[32,36] or the configurations were generated from extensions of ortho- and para- positions[31] but progression of the hydrogenation process was not addressed on flat or curved graphene.

Here, we perform high throughput calculations within an evolutionary framework. We demonstrate formation of lines of hydrogen atoms both on flat and curved graphene sheets. On curved graphene, hydrogenation is predicted to take place as a chain reaction and long lines of hydrogen atoms are shown to be energetically more favourable along the minimum radius of curvature line (Figure 1).

**Methods.** We employ density functional theory based tight-binding (DFTB) method, where we use the DFTB+

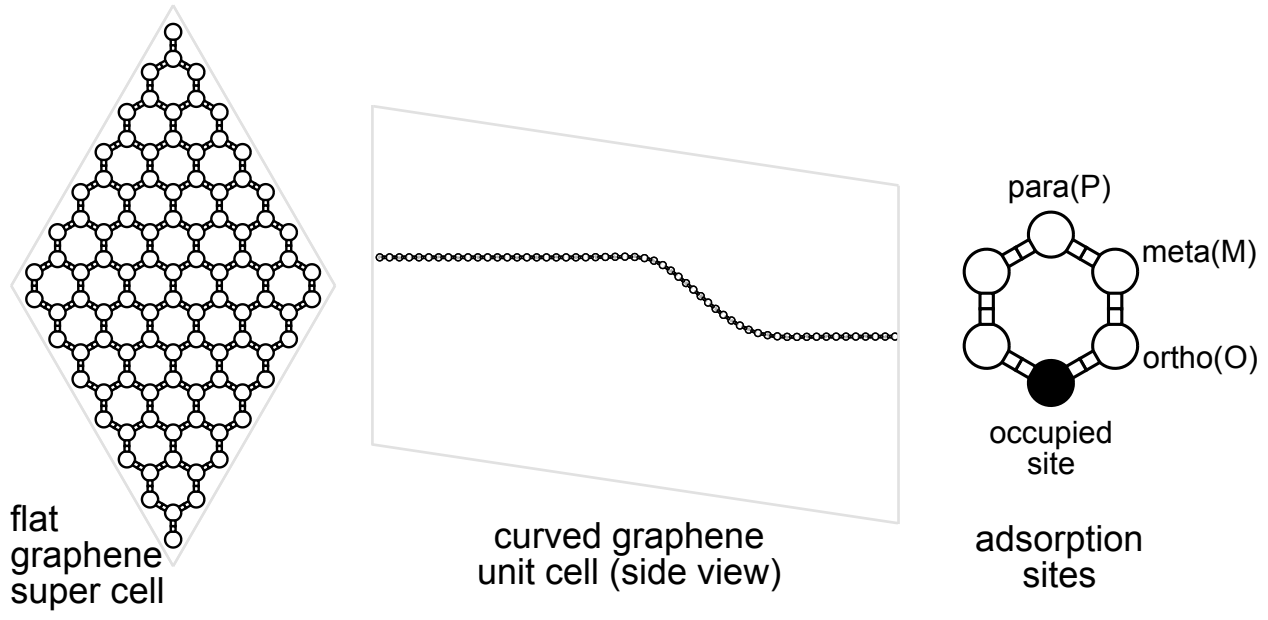

FIG. 2. **Flat and curved graphene** structures are shown and the adsorption sites are labeled as ortho(O)-, para(P)- and meta(M)-configurations with respect to the occupied site.

code [43] and the mio-1-1 parameter set.[44] Density functional theory (DFT) is also used to check the compatibility of the two methods.[45,46] Periodic boundary conditions are employed in all simulations. For flat graphene, a 7×7×1 super cell of the primitive unit cell is used (see Figure 2) with a k-point sampling of 10×10×1 in the Monkhorst-Pack scheme.[47] We fix the super cell lattice parameters corresponding to the carbon-carbon distance of pristine graphene. The vacuum region is set to 10 Å to avoid inter-plane interactions for flat geometry, while it is chosen to be larger than 50 Å for the curved geometries. A unit cell which consists of 144 atoms is built for obtaining the the curved geometry and 1×10×1 super cells of this unit cell are used (see Figure 2) with k-point samplings of 1×3×1 or higher. Structures are optimized until the maximum force component is below the $2\times10^{-2}$ eV/Å. We have checked the strain on graphene super cell after adsorption of hydrogen atoms by re-optimizing the lattice parameters. We have used the most stable hexamer configuration and found that the super cell has shrunk by only 0.35%, which indicates that the adsorption induced strain does not affect our simulations considerably. We disregard spin-polarization in our calculations since the energy difference due to spin polarization is much less than the threshold in the evolutionary algorithm as explained below.

**Flat graphene.** There are several previous studies related to structural properties of hydrogen adsorption on flat graphene[31–38] It was shown that the most preferable positions for hydrogen dimers are the ortho- and para- positions. Ortho(O)-, meta(M)-, and para(P)-positions correspond to the first, second and third nearest neighboring sites, respectively (see Figure 2). Trimer and tetramer adsorption follows extensions of ortho- and para- positions consequently. Before performing high-throughput DFTB simulations, we compare DFTB results against DFT for a relatively large set consisting of 60 configurations, which cover the configurations investigated in the literature.[31,32,36] Details of the comparison are given in the Supplementary Information (Figure S14), which clearly show that DFT and DFTB results are in very good agreement.

We study hydrogen adsorption on flat graphene by using an evolutionary approach, which relies on the selection of favourable configurations by comparing their total energies. (see Figure 3) The algorithm starts with a monomer as the *parent* configuration, from which the *candidate* dimer configurations are generated. The candidate generations include all available adsorption sites up to the fourth nearest neighboring site of the existing occupied sites. The total energies of the relaxed candidate configurations are calculated. The threshold value for the *pre-selection* stage is set to 0.2 eV above the minimum energy of the members of the same *family*, *i.e.* candidates originating from the same parent. Afterwards, a second selection stage is performed, which will be referred to as the *pool-selection*. In pool-selection, the energy threshold is set to 0.2 eV above the global minimum energy of all candidates in the pool. Symmetry is the second criterion during the pool-selection. If there are geometrically equivalent candidates due to symmetries, only one of them is proceeded to the next generation. The successful candidates, *i.e.* stable configurations, are then the parents for the *next generation*.

Starting from dimers, we include a *mutation* mechanism during candidate generation. Mutations alter the configurations of the parent by hopping one of the hydrogens to an available nearest neighboring site. Afterwards, the candidates are generated by adsorption of an hydrogen atom to up to fourth nearest neighboring site, as usual. We simulated all possible mutations on flat graphene from dimers to hexamers, which make 5355 configurations in total. The predicted stable configurations up to tetramers are in exact agreement with literature. Above this size, the evolutionary algorithm finds stable configurations which were not predicted before.

The results of the high-throughput simulations employing the evolution scheme are summarized in Figure 4. Different generations are grouped as rows and depicted with dashed lines. For each generation the number of panels is the sum of the number of successful candidates from previous generation and the number of mutants that could yield stable configurations. The parents are marked with blue squares while the green diamonds show the members of the new generation, i.e the candidates which succeed both selection stages. All candidate configurations are shown in Figure S2-S6 with red circles. If the selection process was carried out only within sister configurations (*i.e.* without pool-selection) the new generation would have a higher number of stable configurations. The configurations that are eliminated during pool-selection are marked with black triangles.

In the first row of Figure 4, one observes that only two candidates, $D_2$ and $D_4$, are selected for the new generation from the parent monomer. They correspond to the ortho- and para-positions. These configurations are selected from 72 candidates (see Figure S2) which are composed of 1 non-mutant and 3 mutant parents. In fact, there are 6 favourable adsorption sites but 4 members $D_3$, $D_5$; $D_1$, $D_6$ are equivalents of $D_2$ and $D_4$, respectively. In agreement with the previous results, the meta-position is found to be unfavourable. The ortho-positioned configuration is more stable than the para-positioned configuration, and there is only 59.2 meV energy difference between them. There is no direct correlation between the dihedral angles and the binding energies, but the C-H bond lengths (1.129 Å, 1.138 Å, 1.142 Å for ortho-, para-, and meta-, respectively) follow the same trend with the binding energies.

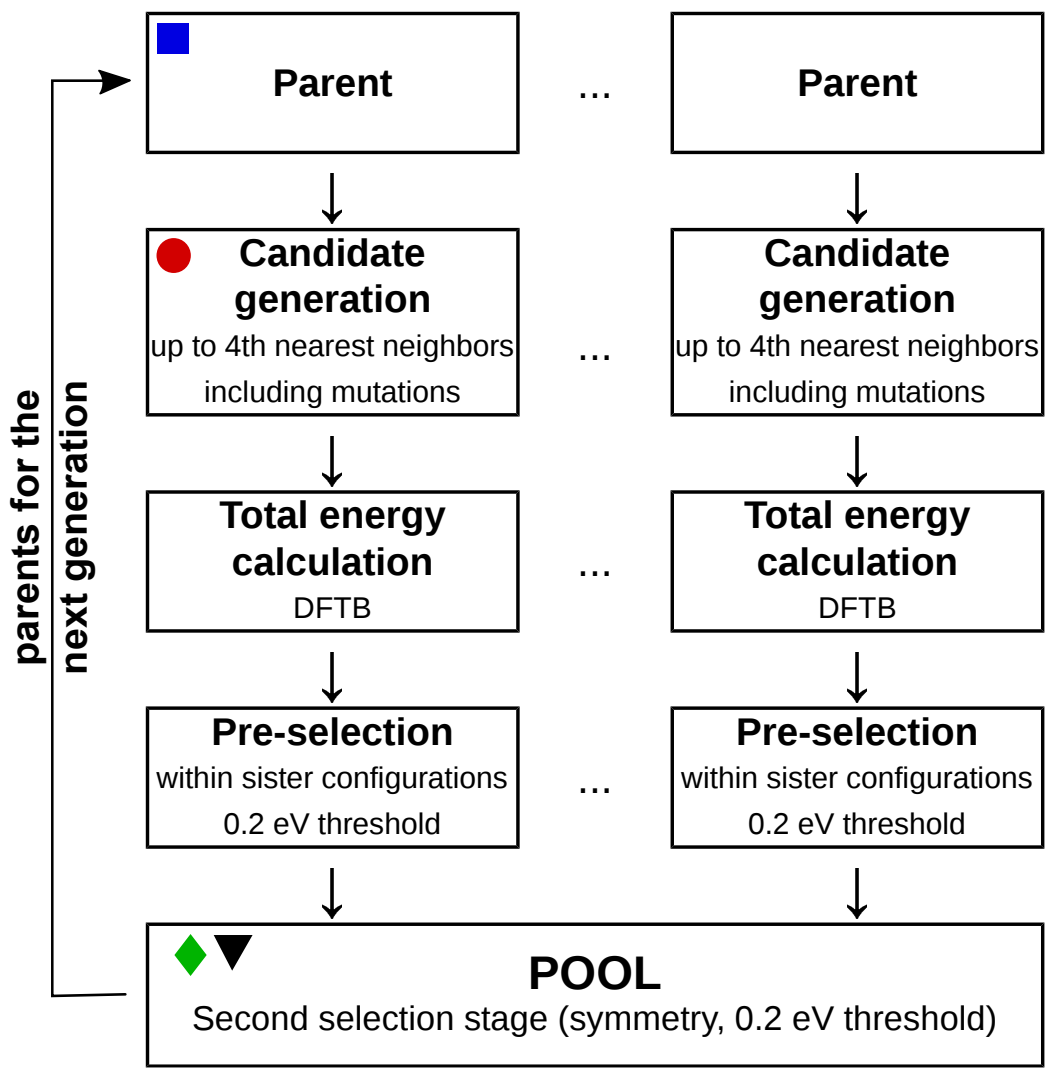

FIG. 3. **Flow chart of the evolutionary algorithm.** The algorithm starts with generation of candidate configurations (red circles) from the parent configurations (blue squares). After DFTB based total energy calculation of the relaxed geometries, pre-selection is performed among the sister configurations, where the threshold energy is set to 0.2 eV. Preselected configurations are collected at a pool, and a second selection procedure is carried out taking total energy and symmetry properties into consideration. Successful candidates (green diamonds) are considered as the stable configurations of the current generation, and they are used as parents for the next generation. Candidates that are eliminated at the pool-selection stage are marked with black triangles. (Blue squares, green diamonds and black triangles are used correspondingly in Figures 4, 6, 7; whereas red circles are used in Figure S2-Figure S8.)

In generating trimers from dimers, there exist 2 non-mutant and 10 mutant parents, which enable 298 candidates (see Figure S3). At the pre-selection stage, 5 successful configurations ($Tr_7$, $Tr_8$, $Tr_9$, $Tr_{10}$, $Tr_{11}$) are found from 2 parents ($D_2$ and $D_4$), one being a mutant ($Tr_9$) (see the third panel in the second row in Fig. 4). The mutation switches a para-configuration to a meta-configuration, which is known to be unstable. It is then stabilized by adsorption of a hydrogen atom to the common ortho-positions of both parent atoms. The stabilization mechanism is in agreement with the literature in the sense that the meta-position is not favourable while the ortho-position is. Two configurations $Tr_{10}$ and $Tr_{11}$) are eliminated due to symmetry. The stable configuration $Tr_8$ occupies P-O positions whereas $Tr_7$ occupies P-P positions.

For tetramers there exist 3 non-mutant and 21 mutant parents which generate 737 candidates (see Figure S4). Among those candidates, we find 4 stable configurations ($Te_{14}$, $Te_{17}$, $Te_{18}$, $Te_{19}$) from 3 parents ($Tr_7$, $Tr_8$, $Tr_9$), one being a mutant ($Te_{14}$) which can be seen in third row of Figure 4. The mutation alters a P-P configuration to a P-M configuration and with the new hydrogen we obtain a P-O-O geometry. It is again confirmed that meta-position is unfavourable. In the tetramer family, five members failed to continue to the next generation at the pool-selection stage (indicated with black triangles). Configuration $Te_{17}$ is composed of P-O-O sites, whereas $Te_{18}$ displays P-O-P geometry. One can make some predictions from the obtained results already. $Te_{17}$ and $Te_{18}$ are precursors of linear chains, whereas a deviation from the linear geometry appears in $Te_{14}$. The evolution of $Te_{19}$ is rather indeterminate. It can either form a hexagonal ring or evolve into a double chain with armchair geometry. The former indicates clustering, while the latter stands for linear patterns.

Next, we generate pentamers from tetramers. There exist 4 non-mutant and 32 mutant parents, which generate 1242 candidates (see Figure S5). Fourth row of Figure 4 shows that the new generation is found to have 7 stable members ($P_{21}$, $P_{22}$, $P_{23}$, $P_{26}$, $P_{28}$, $P_{29}$, $P_{30}$) from 4 parents ($Te_{14}$, $Te_{17}$, $Te_{18}$, $Te_{19}$). The new generation of pentamers does not involve new mutations. All members of the new generation are composed of either only ortho-positioned or mixtures of ortho- and para-positioned hydrogens. There are no O-O-O-O configurations, therefore formation of a hexagonal ring is energetically suppressed at this stage. More interestingly, lines with armchair geometry become the most favourable configurations.

Hexamer configurations are the last step for our calculations on flat graphene. When we consider the results of the hexamer generations, there exist 7 non-mutant and 69 mutant parents which produce 3006 candidates (see Figure S6). From those candidates, the new generation contains 10 non-mutant stable members ($H_{37}$, $H_{44}$, $H_{45}$, $H_{46}$, $H_{47}$, $H_{48}$, $H_{50}$, $H_{52}$, $H_{53}$, $H_{54}$), which can be seen in fifth and the sixth rows of Figure 4. We eliminate 13 of the candidates due to symmetry or because of the ground state coming from a nephew, namely at the pool-selection stage. $H_{37}$ is a mixture of linear and armchair geometries, resembling a broken line. $H_{38}$ is eliminated due to presence of an equivalent, $H_{52}$.

We have checked the strain on graphene super cell after adsorption of six hydrogen atoms by re-optimizing the lattice parameters for the armchair configuration. We find that the super cell has shrunk by only 0.35%, which indicates that the adsorption induced strain does not affect our simulations considerably.

In summary, we generated a pool which is composed of a total of 5355 candidates from dimers to hexamers.

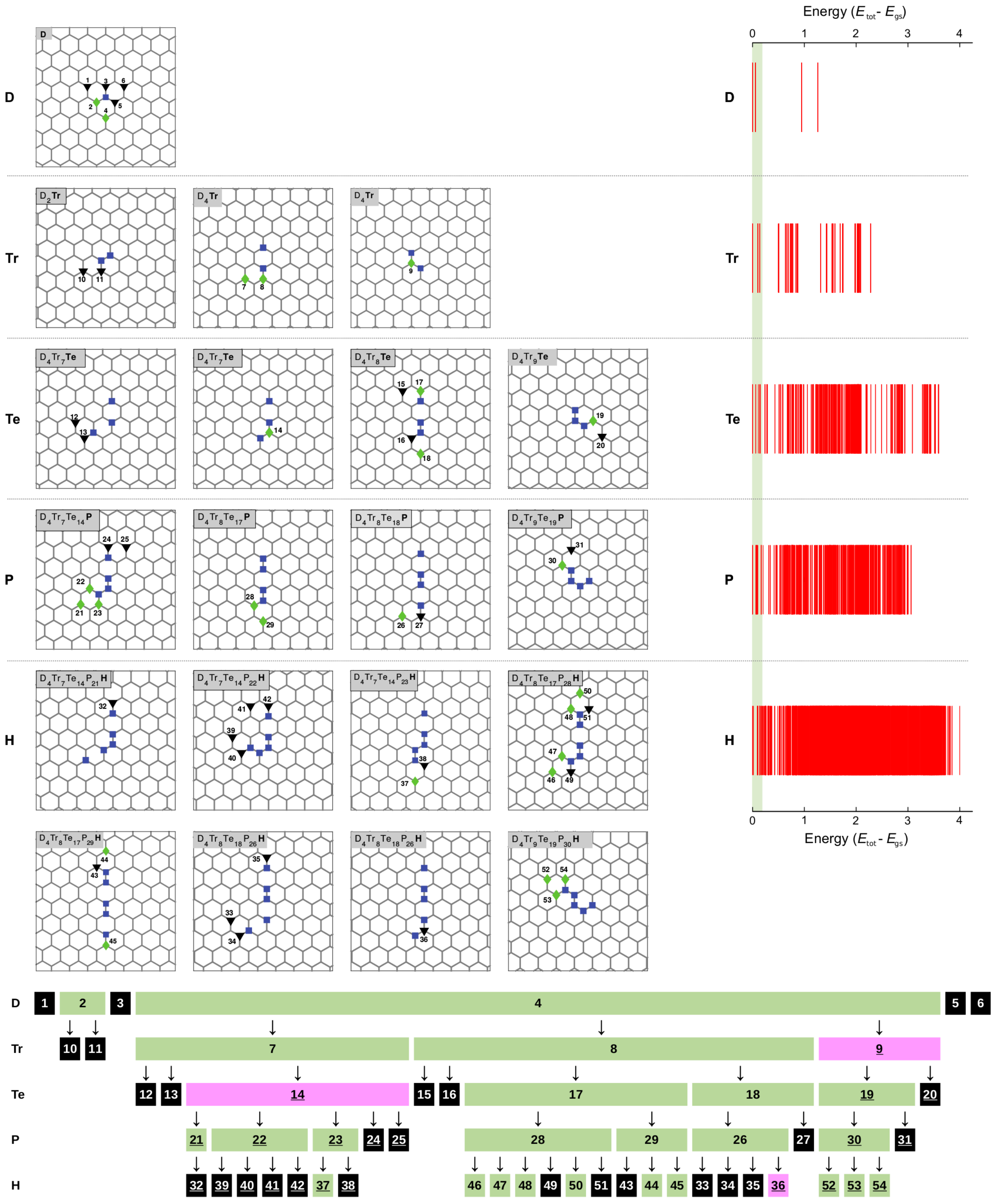

FIG. 4. **Evolution of generations on flat graphene.** Parent atoms (blue squares), stable configurations (green diamonds) and configurations that are eliminated at the pool selection stage (black triangles) are shown throughout the evolution (upper left). Generations are separated with dotted lines and denoted as D, Tr, Te, P, H for dimers, trimers, tetramers, pentamers and hexamers, respectively. Each panel corresponds to a family, the family tree is denoted at the upper left corner of each panel. The evolution of generations can also be tracked at the lower panel, where each row stands for a generation and each box for a candidate with the same enumeration with the geometry plots. The stable, eliminated and mutant configurations are shown with green, black and pink boxes, respectively. Relative total energies of simulated configurations are shown on the right, where the green zone indicates the 200 meV threshold value. The complete list of simulated geometries can be found in Figures S2-S6 and their corresponding energies at Table S-I. See also the Supplementary Animations.

It is clear that ortho- and para- extended combinations are favored, whereas combinations which consist of meta-position are not. The most striking and remarkable result is the alignment of hydrogen atoms in armchair direction and $H_{53}$ is found to be energetically the most preferable configuration. The next most favorable configuration is $H_{45}$, which has a linear alignment with hydrogen atoms occupying A and B sublattices evenly. Energy difference between these two most stable configurations is only 2.67 meV per adsorbant. The six-fold para- configuration (see Conf-2 in Figure S13) is a special case, whose relative total energy is lower than line formations. It does not appear as a stable configuration at the end of the evolution procedure, because its precursor was eliminated at the pool selection stage of tetramers.

Armchair-type line formation was previously shown to be more stable than zigzag-type line formation, as well as formation of triangular and circular clusters,[48] but single lines were not investigated in that work. Single line geometries of tetramers were reported on bilayer graphene but longer lines were not observed.[49] On graphene/SiC(0001) line formations were reported to be as short as a dimer,[50] which are in either in ortho- or para-configurations. Interestingly, scanning tunnelling microscopy images show important electronic contribution of the substrate and the modulation in adsorption energy was reported to be as high as 230 meV. Similarly, Moiré superstructures are known to influence hydrogen adsorption on graphene.[2] Therefore, we speculate that the lack of experimental observation of hydrogen lines on flat graphene may be due to the substrate effects.

**Curved graphene.** A strategy to overcome these effects could be incorporating bending so that substrate induced ripples are overriden by a strong bending and the potential landscape altered by the substrate becomes a minor ingredient compared to the increased chemical reactivity along the minimum radius of curvature line (MRCL). In what follows, we examine hydrogen line formation on curved graphene surfaces.

Armchair direction has the lowest bending modulus[51] and it is also the favoured direction for formation of hydrogen lines. Therefore we consider the substrate direction to be aligned with the armchair direction. The periodic boundary conditions in both directions, parallel and perpendicular to the step edge, are applied. The influence of the interaction with the substrate is included by bending graphene over the step. In the simulations, bending is achieved through constrained relaxation of the atomic positions, where the atoms away from the step edge are fixed and those close the step edge are free during optimization of the atomic forces. The bending angles are chosen to be 52° and 90°, which are determined by considering the bond angles and the strength of substrate graphene interaction. The 52° corresponds to the projection of the tetrahedral angle of $sp^3$ hybridized carbons, and 90° stands for strong interaction between graphene and the substrate. In the absence of hydrogenation, the bending angles are determined solely by the interaction strength. The radius of curvature (RoC) is defined as $R = \left|(1+y')^{3/2}/y''\right|$ and calculated numerically using the interpolated curve. The lengths of the curved parts across the step are chosen so as to remove the tensile stress after hydrogenation. For different interaction strengths, which correspond to different substrates, RoC is found to be in the range between 13.5 Å and 3.3 Å, where the step height is kept as 10 Å.[28,30]. In Figure 5(b,d) the positions of carbon atoms (red circles) and the fitted curve (red curve) are plotted together with the calculated RoC (blue) and its inverse (green). In Figure 5(a,c), the total energies are plotted with reference to the minimum energy configurations as the adsorption site is varied for both bending angles. Top and side views of tested adsorption sites are indicated from 1 to 11 in the insets. The reactivity of graphene increases with reduced RoC, and the total energy increases almost symmetrically with increasing RoC, where the minimum energy is achieved at the site with the smallest RoC. In 52°-bending the minimum RoC is 16 Å, while in 90° bending we find the minimum RoC close to 10 Å, which are close to the previous results.[30] We also perform single hydrogen adsorption calculations on curved graphene for 90°-bending by using SIESTA. All trends in total energy

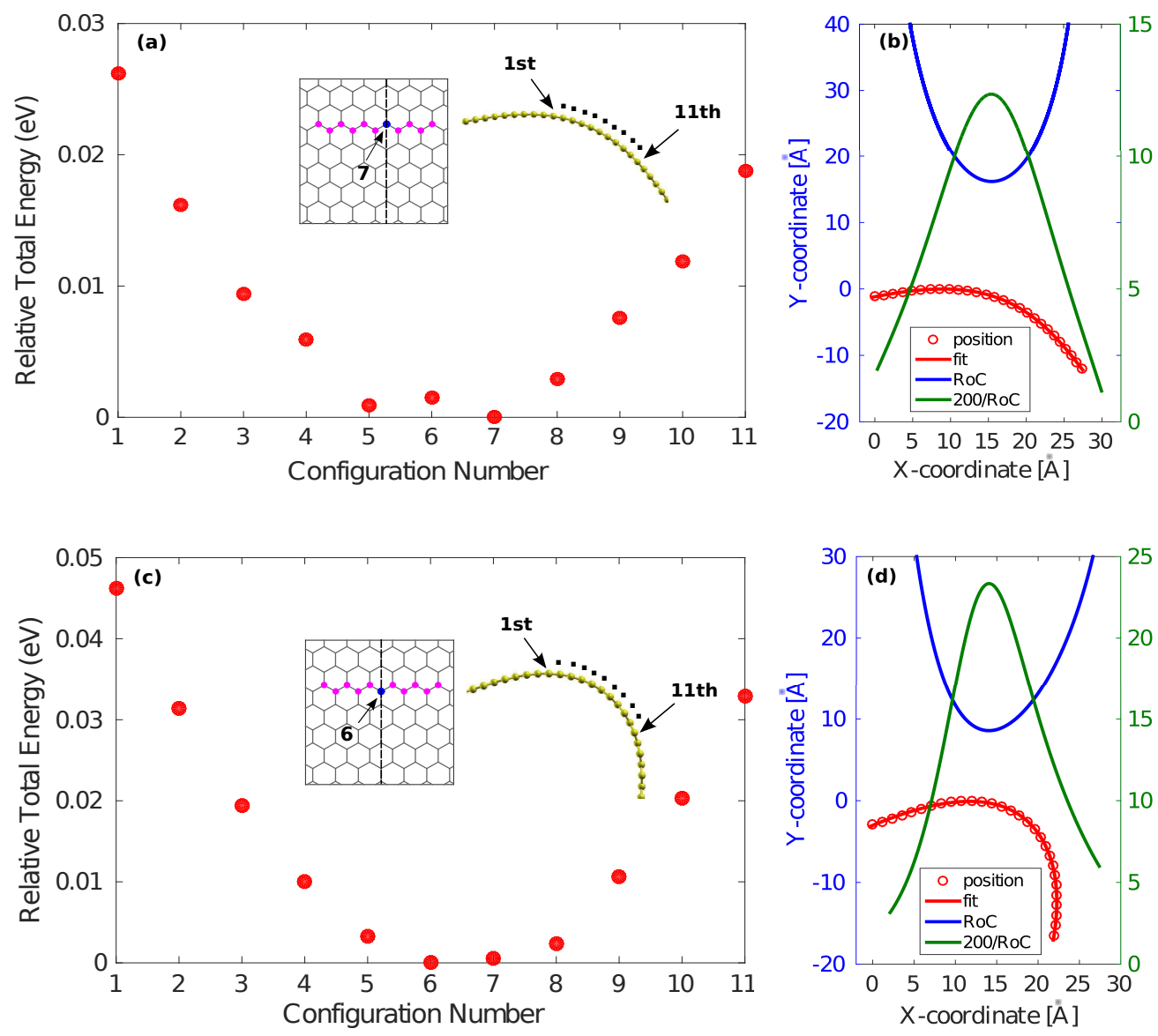

FIG. 5. **Curvature and binding to curved graphene.** Energetics of single hydrogen adsorption on curved graphene with bending angle 52° and 90°, (a and c, respectively). In the insets top-view and side-view of the configurations are illustrated. Pink dots denote tested-configurations and blue dots show most favourable positions. Configurations-7 and 6 ensure the positions of the minimum radius of curvature lines (MRCL) for 52°- and 90°-bending. The MRCLs are depicted in the insets. In (b) and (d) red circles indicate graphene atoms, the blue lines show the radius of curvature(RoC) and the green lines show the inverse radius of curvature (IRoC).

from DFT are reproduced by DFTB (see Figure S14 and Figure S15). Binding energies are calculated using

$$E^{(n)}_{\text{binding}} = (E_{\text{graphene}} + nE_{\text{H}} - E_{\text{graphene+H}_\text{n}})/n. \quad (1)$$

The binding energies of single hydrogen on 90°-bent graphene are larger than those of the 52°-bent graphene due to the curvature effect (see in Table S-XIII and Table S-XVI for 52°-bending; Table S-XVIII and Table S-XXI for 90°-bending).

The number of atoms in the simulation cell is significantly larger for curved graphene. On top of that, the number of possible configurations is multiplied because of the broken symmetries due to bending. These make it impossible to simulate all possible configurations. Therefore, equipped with the information from the evolution on flat graphene, we reduce the number of candidates significantly on curved graphene by focusing on the formation of lines of hydrogen atoms.

**52°-Bending.** In Figure 5(a), it is shown that configuration-7 is the most favourable adsorption site for hydrogen on curved graphene (52°) which coincides with the MRCL. Configuration-7 is taken as the parent for the dimer generation. Successful candidates as well as those eliminated during the pool selection are shown in Figure 6, while all tested candidates on 52°-bent graphene are presented in Figure S7. We note that, enumeration of configurations in Figure S7 is independent from the enumeration in Figure 6.

For dimers, we consider 21 candidates, which are distributed equally on the left and the right hand sides of the MRCL (see first row of Figure S7). At the pre-selection stage, six candidates ($D_1$...$D_6$) are found to be stable. (Figure 6) When selection process is considered, dimers (52°) produced more parents for the next generation than flat graphene (0°) due to the symmetry breaking with the curvature. However these parents are ortho- and para-positioned hydrogens as in flat graphene. Only two candidates $D_1$ and $D_5$ are eliminated during the pool selection. Relative total energies and binding energies per hydrogen atom of 21 candidates are summarized in Table S-II and Table S-XII, respectively. The most favourable dimer configuration is found to be $D_4$ on 52°-bent graphene.

For trimers, 43 candidates are generated from 4 parents (see the second row of Figure S7). Only 4 of them succeed after the selection process. (see Figure 6) For all trimer configurations considered, P-O positioned $Tr_{11}$ and P-O positioned $Tr_{14}$ making a 60°-angle with the MRCL are found to be the most stable configurations. It is interesting that no P-P positioned configurations appear in 52°-bending whereas P-P positioned $D_4Tr_7$ (0°) is one of the most favourable configuration. It is expected that a P-P positioned configuration would be generated from $D_3Tr$ (52°) but two of the parent's atoms are located on the MRCL and this causes an increase in the energy cost to adsorb a third hydrogen at the para-position. Relative total energies and binding energies per hydrogen are summarized in Table S-III and Table S-XIII, respectively.

Tetramer generation consists of 94 candidates from 4 parents (see the third row of Figure S7). Seven candidates are selected as parents for the next generation. Tetramer configurations that are stable on flat graphene are favored on 52°-bent case, as well (Figure 6). One should note that, the stable configurations are etiher mixtures of ortho- and para- positions or pure ortho positioned configurations. In addition, pure P-P-P positioned configurations are eliminated at pre-selection stage in 52°-bending while those of flat graphene are eliminated at pool-selection stage. Another point is O-O-P positioned $Te_{16}$ tends to form a single line on the second-MRCL, the parallel line on the left or on the right of the MRCL. In family $D_2Tr_9Te$, $Te_{20}$ is kept in order to observe the effect of a shift in the position parallel to the MRCL in the subsequent generation. It can be seen from Table S-IV and Table S-XIV that $Te_{23}$ ($Te_3$ in Figure S7), which supports linear growth on the MRCL, is the most favorable configuration.

In generating pentamers from tetramers, 108 candidates have been analyzed (see the fourth and the fifth rows of Figure S7). These candidates are derived from 7 parents, thus we examine 7 families in this generation. The first significant difference in terms of the number of families between flat and curved graphene is realized in pentamers. $P_{22}$ (0°) was selected for next generation but the corresponding geometry $P_{33}$ (52°) is failed at the pool-selection stage. Linear hydrogen chain formation on the first- and the second-MRCL become definite with the configurations $P_{42}$, $P_{40}$ and $P_{28}$. In addition, armchair geometry reappears with $P_{31}$ and $P_{37}$. It is interesting to note that although $P_{30}$ is closer to the MRCL than $P_{38}$, $P_{30}$ is eliminated. The reason can be that most of the parents on the MRCL lift carbon atoms and the formation of kink on MRCL makes it easier to adsorb a hydrogen atom. That is, the reaction barrier is lowered, giving rise to a mechanical chain reaction. The same configuration on flat graphene $P_{21}$ was also transferred to the next generation. There are two families $D_2Tr_8Te_{18}P$ and $D_6Tr_{14}Te_{27}P$ in pentamers, which can not produce any successful candidates. Nevertheless, we proceed them to the next generation in order to check the growth of lines across the MRCL. In tetramers, it was possible to have linear configurations making 60°-angle with the MRCL but in pentamers those formations are all eliminated and only lines along the MRCL are favored, a direct consequence of curvature. In 90°-bending, this effect becomes more pronounced, as it will be discussed below.

Finally, in generating hexamers from pentamers, 195 candidates from 9 parents are investigated (see the sixth and the seventh rows of Figure S7). The second criterion of the pool-selection, which is related to the symmetry properties, is not taken into account in order to display all possible preferential configurations. $H_{53}$, $H_{54}$ and $H_{70}$ (Figure 6) are eliminated in curved graphene 52°-bending but their corresponding configurations $H_{47}$ and $H_{46}$ in flat graphene are stable, which indicates that growth of single hydrogen lines in directions except the MRCL di-rection are suppressed. In summary, the most stable configurations have armchair pattern ($H_{58}$ and $H_{64}$). Both evolve from $D_2$ but following different paths. In addition, single line formation is also favored. It is preferable when there are no unpaired hydrogens, while armchair orientation has the minimum energy even with unpaired hydrogens.

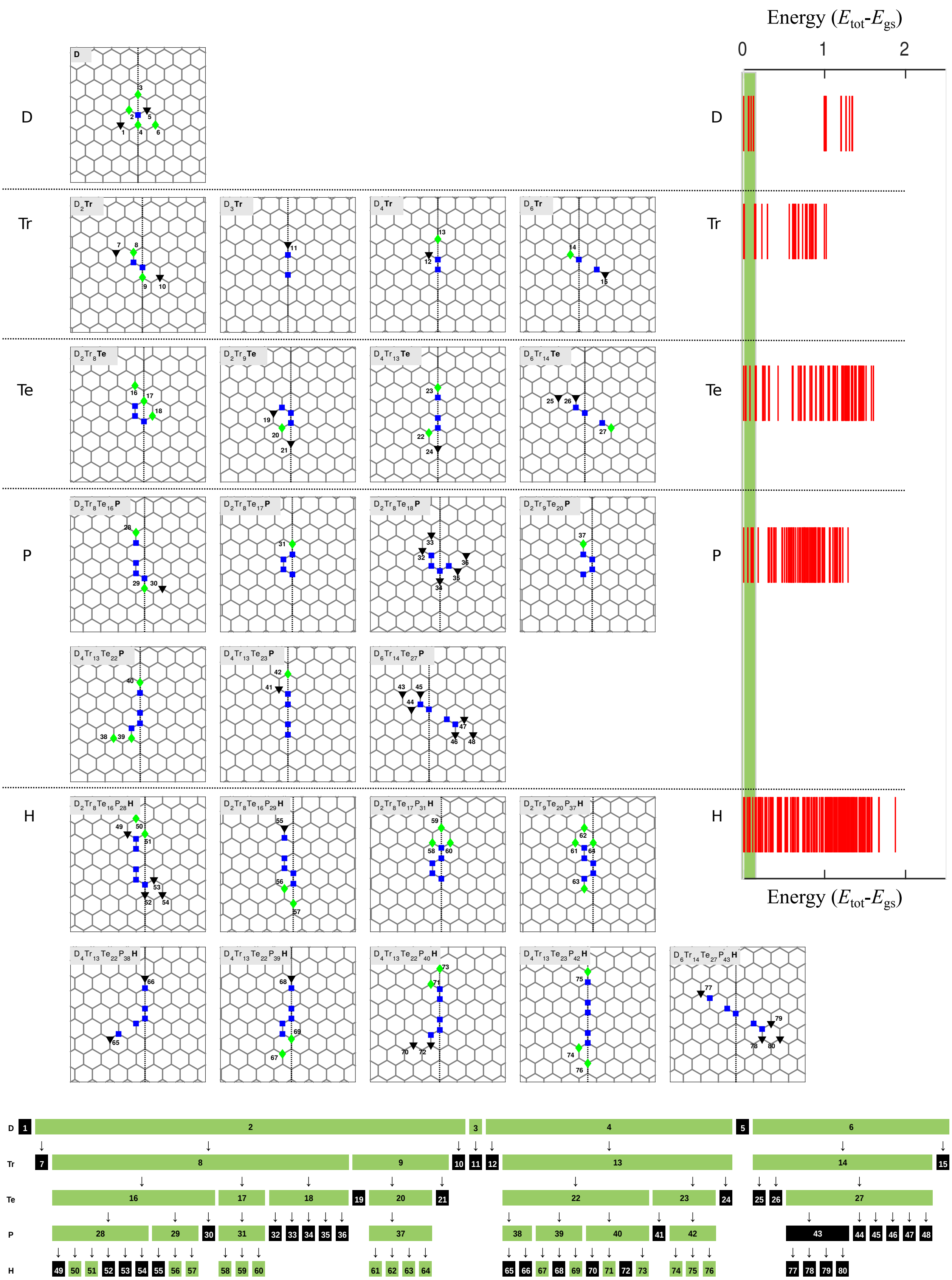

FIG. 6. **Evolution of generations during the hydrogenation of curved graphene (52°).** The same color codes and notation are applied as in Figure 4. The complete list simulated structures can be found in Figure S7.

In order to examine directed growth for larger numbers of hydrogen atoms, we consider single-line and armchair-type configurations of 10 atoms. When hydrogen atoms lie along the MRCL, binding energy per hydrogen atom is enhanced by 116 meV and 131 meV compared to those lie across the MRCL with 60°-angle for single line and armchair geometries, respectively. For 52° bending, the maximum length of stable single lines is as short as a tetramer.

**90°-Bending**. Relative total energies and the binding energies for 90°-bent graphene are listed in Tables S-VII-S-XI and Tables S-XVII-S-XXI. In Figure 5(c), the most stable monomer, namely the parent of the dimers marks the position of the MRCL for 90°-bent graphene. The family tree is shown in Figure 7) whereas a complete list of tested configurations can be found in Figure S8.

The first difference between 52°- and 90°-bending appears in the trimer generation. $Tr_{14}$ is suppressed due to the stronger curvature effect in 90°-bending, which succeeded in 52°. This means that only the lines as short as a hydrogen dimer are stable if they are not aligned with the MRCL. The family $D_6Tr(90°)$ can not produce any candidates that can succeed the pool-selection. However we proceed this family in order to check the energetics of lines across the MRCL. Family $D_2Tr_8Te_{18}P(52°)$ is not observed in 90°-bending, because its parent was eliminated in the previous generation. Family $D_4Tr_{13}Te_{23}P(90°)$ is similar to $D_4Tr_{13}Te_{22}P(52°)$ except the configuration $P_{38}$, which is also eliminated due to stronger curvature. The differences between 52°- and 90°-bending become more clear in hexamers. Even though $D_2Tr_8Te_{16}P_{28}H(52°)$ and $D_2Tr_8Te_{16}P_{29}H(90°)$ originate from the same parents, higher curvature does not allow succession of the configurations which are not on the MRCL. In 52°-bending, hydrogens can prefer to be ordered along a short line across the MRCL within the same family, whereas 90°-bending does not allow such a geometry. $H_{52}(52°)$ which was eliminated, appears as a successful candidate as $H_{43}(90°)$. This geometrical change reveals that single line formation on the second-MRCL does not continue in 90° bending. In the family $D_2Tr_8Te_{16}P_{30}H$, $H_{44}$ which is the same as $H_{55}(52°)$, creates a single line on the second-MRCL. Both $H_{58}$ and $H_{59}$ form a single line on the MRCL.$H_{59}$ leaves an unoccupied site for this reason, $H_{59}$ is less favoured in total energy when compared to $H_{58}$. $H_{57}(90°)$ is eliminated during the pool selection. However, $H_{74}(52°)$ which is identical with $H_{57}(90°)$ succeeds through the whole selection process. This difference is originating from the the kink formation being stronger for 90°-bending than it is in 52°-bending. In summary, the main difference between 52° and 90°-bending is in the length of hydrogen lines if they are not aligned with the MRCL. The number of eliminated candidates, especially in hexamer generation, are less than those of 52°-bending due to the decrease in RoC. This result indicates that families of pentamer and hexamer generations are more stable than in 52° bending. As it is the case for 52°-bent graphene, armchair geometry is more favourable than the linear geometry along the MRCL. Binding energy per hydrogen atom in the armchair configuration is about 40 meV more than that of linear configuration. Growth of single line and armchair configurations consisting of 10 hydrogen atoms are compared along and across the MRCL and it is found that the binding energies per adsorbate are enhanced by 200 meV and 223 meV for single line and armchair geometries, respectively.

Considering lines consisting of 6 hydrogen atoms, bending increases the binding energies by 151 meV (176 meV) and 265 meV (289 meV) per hydrogen atom fro 52°- (90°-)bending with single line and armchair configurations, respectively. This is a clear indication that curvature favours directed growth of hydrogen lines and it may overcome substrate effects.

**Conclusion**. High throughput simulations show that our evolutionary algorithm is able to predict more stable configurations than those studied before. It is more preferable for hydrogen to adsorb on lines rather than making clusters, which obey the symmetries of the hexagonal lattice. Moreover, the line formation has a preferred crystallographic orientation, namely the armchair direction, while growth along the zigzag direction is suppressed. Combined with the effect of bending on chemical reactivity, the selection process eliminates hydrogen lines which are not aligned with the MRCL of bent graphene and a directed growth becomes possible. At intermediate bending angles, line formations crossing the MRCL are possible up to tetramers. When the RoC is smaller, the length shortens down to a hydrogen dimer. The growth can be viewed as a mechanical chain reaction. The reaction barrier is lowered by both bending of the surface and proximity to an occupied site. As a result kinked graphene can be fabricated, where the electrons on the opposite sides of the hydrogen line are decoupled.[29]

---

* haldunsevincli@iyte.edu.tr

[1] Y. Lu and Y. P. Feng, Nanoscale **3**, 2444 (2011).
[2] R. Balog, B. Jorgensen, L. Nilsson, M. Andersen, E. Rienks, M. Bianchi, M. Fanetti, E. Laegsgaard, A. Baraldi, S. Lizzit, Z. Sljivancanin, F. Besenbacher, B. Hammer, G. P. Thomas, P. Hofmann, and

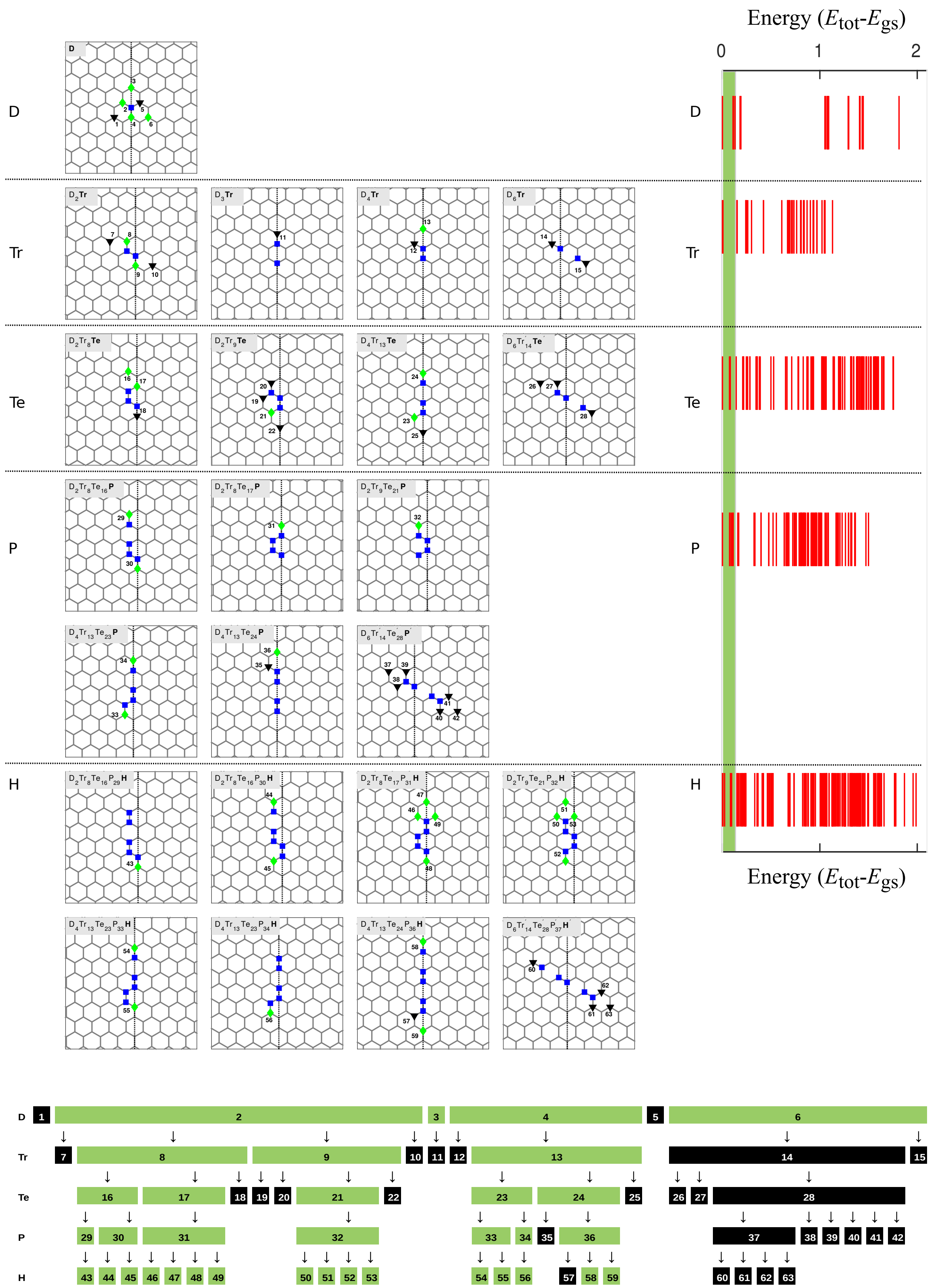

FIG. 7. **Evolution of generations during the hydrogenation of curved graphene (90°).** The same color codes and notation are applied as in Figure 4. The complete list simulated structures can be found in Figure S8.

L. Hornekaer, Nat Mater **9**, 315–319 (2010).
[3] C. Lin, Y. Feng, Y. Xiao, M. Dürr, X. Huang, X. Xu, R. Zhao, E. Wang, X.-Z. Li, and Z. Hu, Nano Letters **15**, 903 (2015), pMID: 25621539.
[4] D.-S. Zhang and J. Li, Physics Letters A **378**, 68 (2014).
[5] D. Haberer, D. V. Vyalikh, S. Taioli, B. Dora, M. Farjam, J. Fink, D. Marchenko, T. Pichler, K. Ziegler, S. Simonucci, M. S. Dresselhaus, M. Knupfer, B. Büchner, and A. Grüneis, Nano Letters **10**, 3360 (2010), pMID: 20695447.
[6] J. O. Sofo, A. S. Chaudhari, and G. D. Barber, Phys. Rev. B **75**, 153401 (2007).
[7] P. Chandrachud, B. S. Pujari, S. Haldar, B. Sanyal, and D. G. Kanhere, Journal of Physics: Condensed Matter **22**, 465502 (2010).
[8] D. C. Elias, R. R. Nair, T. M. G. Mohiuddin, S. V. Morozov, P. Blake, M. P. Halsall, A. C. Ferrari, D. W. Boukhvalov, M. I. Katsnelson, A. K. Geim, and K. S. Novoselov, Science **323**, 610 (2009).
[9] B. S. Pujari, S. Gusarov, M. Brett, and A. Kovalenko, Phys. Rev. B **84**, 041402 (2011).
[10] Z. F. Wang, Y. Zhang, and F. Liu, Phys. Rev. B **83**, 041403 (2011).
[11] W. Bao, F. Miao, Z. Chen, H. Zhang, W. Jang, C. Dames, and C. N. Lau, Nat Nano **4**, 562 (2009).
[12] S.-Y. Lin, S.-L. Chang, F.-L. Shyu, J.-M. Lu, and M.-F. Lin, Carbon **86**, 207 (2015).
[13] L. A. Chernozatonskii, P. B. Sorokin, and J. W. Brüning, Applied Physics Letters **91**, 183103 (2007).
[14] L. A. Chernozatonskii and P. B. Sorokin, The Journal of Physical Chemistry C **114**, 3225 (2010).
[15] V. V. Shunaev and O. E. Glukhova, The Journal of Physical Chemistry C **120**, 4145 (2016).
[16] O. Glukhova and M. Slepchenkov, Nanoscale **4**, 3335 (2012).
[17] R. Balog, B. Jørgensen, J. Wells, E. Lægsgaard, P. Hofmann, F. Besenbacher, and L. Hornekær, Journal of the American Chemical Society **131**, 8744 (2009).
[18] A. Rossi, S. Piccinin, V. Pellegrini, S. de Gironcoli, and V. Tozzini, The Journal of Physical Chemistry C **119**, 7900 (2015).
[19] S. Park, D. Srivastava, and K. Cho, Nano Letters **3**, 1273 (2003).
[20] P. Ruffieux, O. Gröning, M. Bielmann, P. Mauron, L. Schlapbach, and P. Gröning, Phys. Rev. B **66**, 245416 (2002).
[21] G. Sclauzero and A. Pasquarello, Phys. Rev. B **85**, 161405 (2012).
[22] T. Cavallucci and V. Tozzini, The Journal of Physical Chemistry C **120**, 7670 (2016).
[23] V. Tozzini and V. Pellegrini, The Journal of Physical Chemistry C **115**, 25523 (2011).
[24] S. Goler, C. Coletti, V. Tozzini, V. Piazza, T. Mashoff, F. Beltram, V. Pellegrini, and S. Heun, The Journal of Physical Chemistry C **117**, 11506 (2013).
[25] V. D. Camiola, R. Farchioni, t. Cavallucci, A. Rossi, V. Pellegrini, and V. Tozzini, Frontiers in Materials **2** (2015).
[26] D. W. Boukhvalov, M. I. Katsnelson, and A. I. Lichtenstein, Phys. Rev. B **77**, 035427 (2008).
[27] F. Mercuri, International Journal of Hydrogen Energy (2017).
[28] T. Low, V. Perebeinos, J. Tersoff, and P. Avouris, Phys. Rev. Lett. **108**, 096601 (2012).
[29] J. T. Rasmussen, T. Gunst, P. Bøggild, A.-P. Jauho, and M. Brandbyge, Beilstein Journal of Nanotechnology **4**, 103 (2013).
[30] H. Sevinçli and M. Brandbyge, Applied Physics Letters **105**, 153108 (2014).
[31] Z. Sljivancanin, M. Andersen, L. Hornekær, and B. Hammer, Phys. Rev. B **83**, 205426 (2011).
[32] Z. Sljivancanin, E. Rauls, L. Hornekaer, W. Xu, F. Besenbacher, and B. Hammer, The Journal of Chemical Physics **131**, 084706 (2009).
[33] Y. Ferro, D. Teillet-Billy, N. Rougeau, V. Sidis, S. Morisset, and A. Allouche, Phys. Rev. B **78**, 085417 (2008).
[34] M. Moaied, J. A. Moreno, M. J. Caturla, F. Ynduráin, and J. J. Palacios, Phys. Rev. B **91**, 155419 (2015).
[35] S. Casolo, O. M. Løvvik, R. Martinazzo, and G. F. Tantardini, The Journal of Chemical Physics **130**, 054704 (2009).
[36] T. Fei Cao, L. Feng Huang, X. Hong Zheng, P. Lai Gong, and Z. Zeng, Journal of Applied Physics **113**, 173707 (2013).
[37] N. Rougeau, D. Teillet-Billy, and V. Sidis, Chemical Physics Letters **431**, 135 (2006).
[38] L. Hornekær, E. Rauls, W. Xu, i. c. v. Šljivančanin, R. Otero, I. Stensgaard, E. Lægsgaard, B. Hammer, and F. Besenbacher, Phys. Rev. Lett. **97**, 186102 (2006).
[39] i. c. v. Šljivančanin, Phys. Rev. B **84**, 085421 (2011).
[40] M. S. Alam, F. Muttaqien, A. Setiadi, and M. Saito, Journal of the Physical Society of Japan **82**, 044702 (2013).
[41] M. Andersen, L. Hornekær, and B. Hammer, Phys. Rev. B **86**, 085405 (2012).
[42] R. Balog, M. Andersen, B. Jørgensen, Z. Sljivancanin, B. Hammer, A. Baraldi, R. Larciprete, P. Hofmann, L. Hornekær, and S. Lizzit, ACS Nano **7**, 3823 (2013), pMID: 23586740.
[43] B. Aradi, B. Hourahine, and T. Frauenheim, The Journal of Physical Chemistry A **111**, 5678 (2007), pMID: 17567110.
[44] M. Elstner, D. Porezag, G. Jungnickel, J. Elsner, M. Haugk, T. Frauenheim, S. Suhai, and G. Seifert, Phys. Rev. B **58**, 7260 (1998).
[45] P. Hohenberg and W. Kohn, Phys. Rev. **136**, B864 (1964).
[46] W. Kohn and L. J. Sham, Phys. Rev. **140**, A1133 (1965).
[47] H. J. Monkhorst and J. D. Pack, Phys. Rev. B **13**, 5188 (1976).
[48] M. Li, L. Wang, N. Yu, X. Sun, T. Hou, and Y. Li, J. Mater. Chem. C **3**, 3645 (2015).
[49] T. F. Cao, L. F. Huang, X. H. Zheng, W. H. Zhou, and Z. Zeng, The Journal of Chemical Physics **139**, 194708 (2013).
[50] P. Merino, M. Švec, J. I. Martínez, P. Mutombo, C. Gonzalez, J. A. Martín-Gago, P. L. de Andres, and P. Jelinek, Langmuir **31**, 233 (2015).
[51] Q. Lu, M. Arroyo, and R. Huang, Journal of Physics D: Applied Physics **42**, 102002 (2009).
[52] J. M. Soler, E. Artacho, J. D. Gale, A. García, J. Junquera, P. Ordejón, and D. Sánchez-Portal, Journal of Physics: Condensed Matter **14**, 2745 (2002).
[53] N. Troullier and J. L. Martins, Phys. Rev. B **43**, 1993 (1991).

# Supplementary Video

## Directed growth of hydrogen lines on graphene: high-throughput simulations powered by evolutionary algorithm

The Supplementary Video can be reached at
`http://hsevinclilab.iyte.edu.tr/DirectedGrowth/evolution.avi`

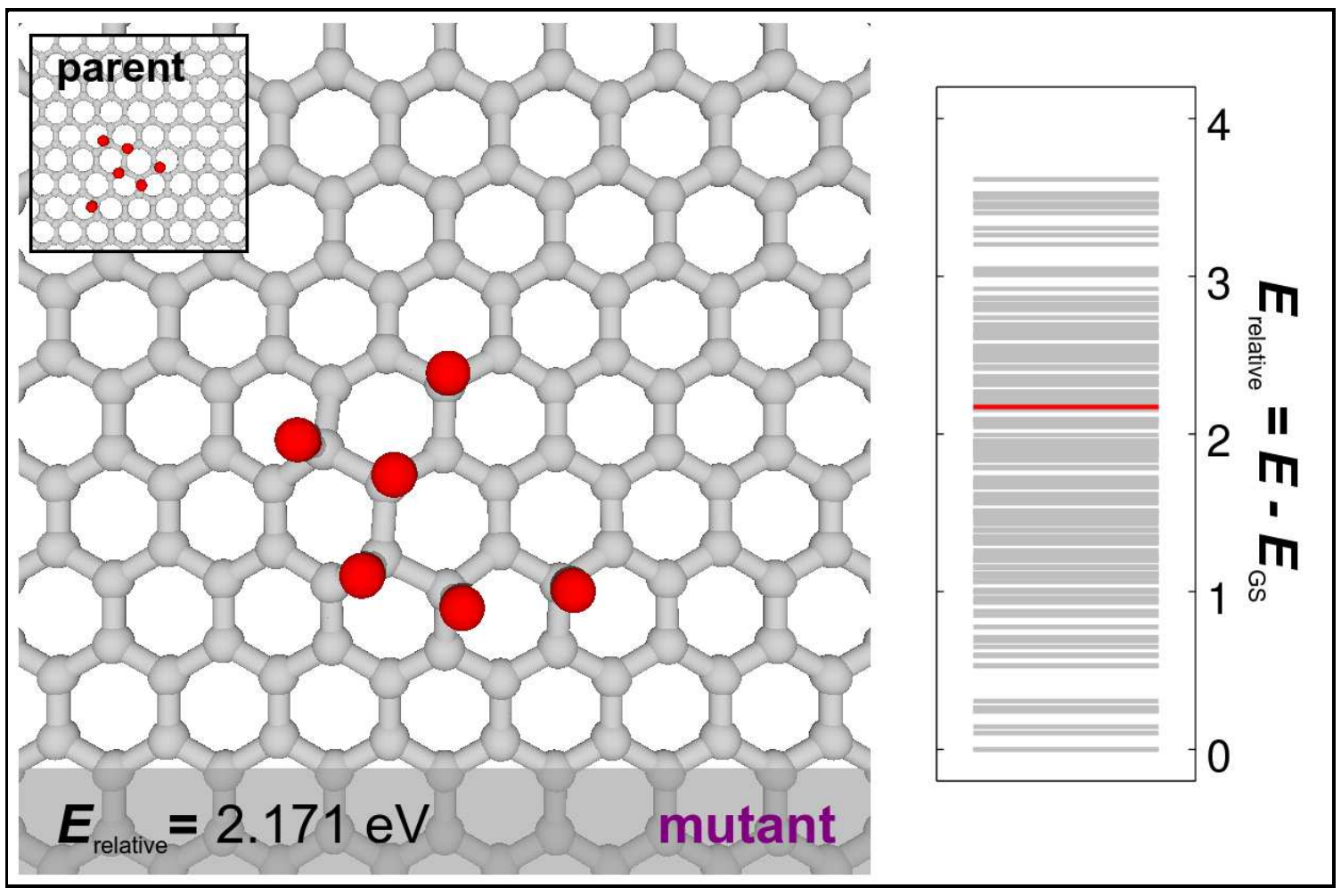

FIG. SV1. Snapshot from Supplementary Video.

The video contains information on the evolution of hydrogen lines on flat graphene and following the flowchart of the evolutionary algorithm shown in Figure 3. Starting from dimers, each candidate from each parent are shown in the left panel. (see Figure SV1) Its parent is shown on the upper left frame and its energy relative to the lowest energy configuration is given at the bottom, $E_{\rm relative} = E - E_{\rm GS}$. Mutant configurations are marked with a purple indicator at the bottom. The right panel is reserved for the energy spectrum of the candidates. The gray lines show $E_{\rm relative}$ of the candidates that have been considered so far and red line is indicates $E_{\rm relative}$ of the current candidate.

# Supplementary Information

## Directed growth of hydrogen lines on graphene: high-throughput simulations powered by evolutionary algorithm

### CONTENTS

### I. SYSTEM SETUP

Simulations regarding hydrogenation of flat graphene are performed on 7×7×1 super cells of the primitive unit cell using periodic boundary conditions in the plane of graphene and a 10 Å vacuum to avoid interlayer interactions. (Figure S1-a) For curved graphene, the simulation cell is chosen so as to allow periodic boundary conditions. (Figure S1-c) Straight parts of the layer are kept fixed to maintain desired curvature. Ortho-, meta- and para positions on a hexagon are defined as the first, the second and the third nearest neighbors with respect to the reference atom. (Figure S1-b)

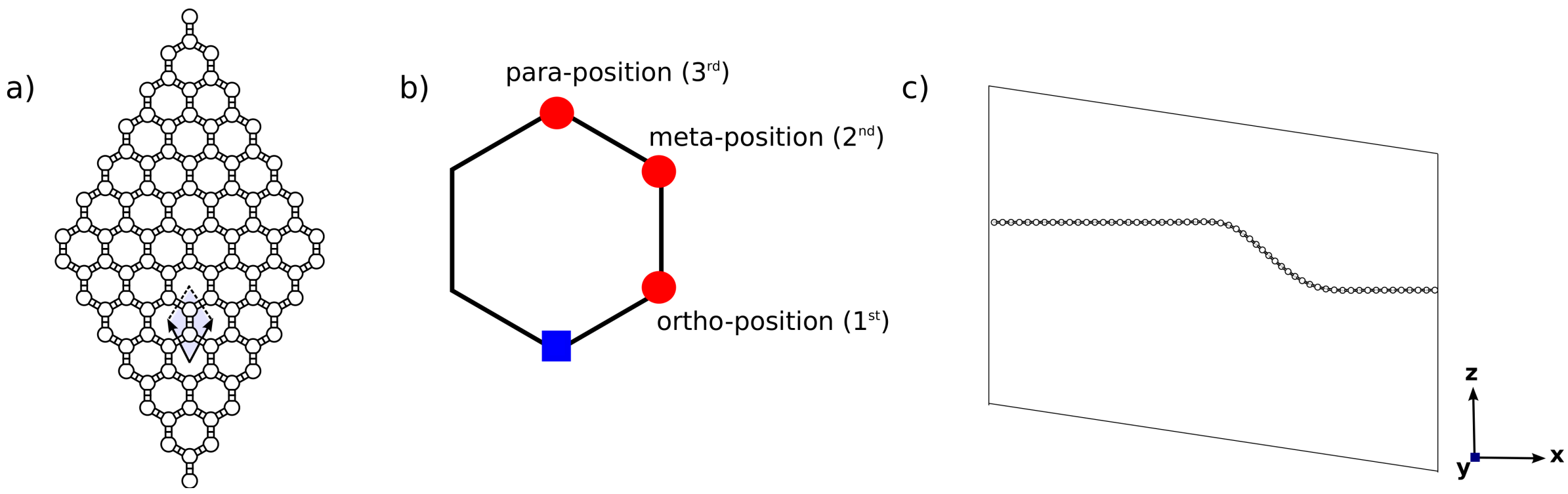

FIG. S1. (a) Primitive unit cell and the 7×7×1 super cell (7x7x1) of flat graphene. (b) Hydrogen adsorption sites: Ortho-, meta-, and para- positions. (c) Illustration of the curved graphene unitcell with 52°-bending.

## II. HIGH-THROUGHPUT CALCULATIONS

### A. High Throughput Calculations on Flat Graphene

We start with a monomer, from which 1 non-mutant and 3 mutant parents are generated. (see Figure S2) Mutant parents are generated by shifting the initial hydrogen to its three first nearest neighbouring sites. In fact, all 4 families are equivalent in this first stage of evolution. Total energy calculations are performed for 72 candidates in dimer generation. In trimers there exist 2 non-mutant and 10 mutant parents which enable 298 candidates. 2 non-mutant parents are shown in the first panel of the first row and the last panel of the second row in Figure S3. In tetramers there exist 3 non-mutant and 21 mutant parents which generate 737 candidates. 3 non-mutant parents are shown in the first panel of the first row, the third panel of the second row and the first panel of the fifth row in Figure S4. In pentamers, there exist 4 non-mutant and 32 mutant parents, which generate 1242 candidates. 4 non-mutant parents are shown in the first panel of the first row, the last panel of the second row, the third panel of the fifth row and the last panel of the seventh row in Figure S5. In hexamers there exist 7 non-mutant and 69 mutant parents which produce 3006 candidates. 7 non-mutant parents are shown in the first panel of the first row, the first panel of the third row, the third panel of the sixth row, the third panel of the nineth row, the first panel of the twelfth row, the third panel of the fourteenth row, the third panel of the seventeenth row in Figure S6

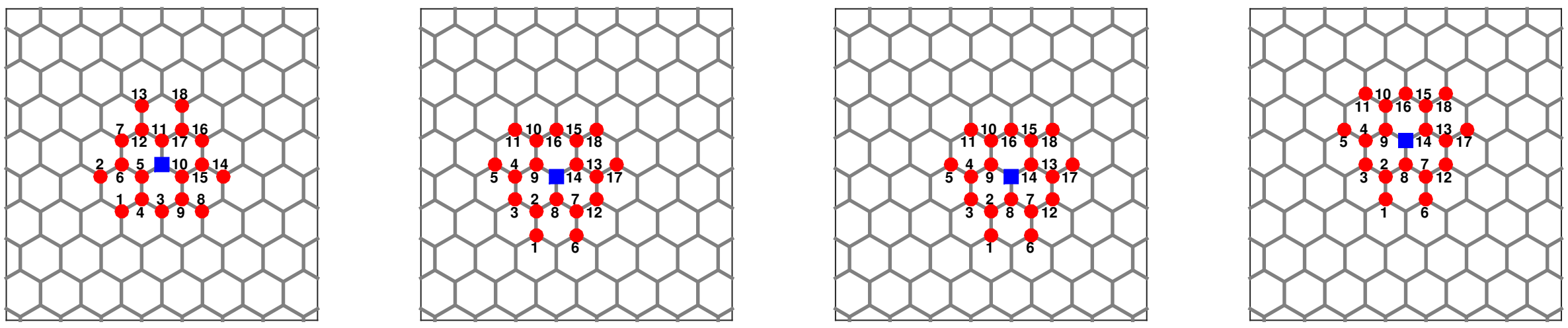

FIG. S2. 1 non-mutant and 3 mutant families of Dimer generation in flat graphene. 4 families consist of 72 candidates are generated from 1 parent.

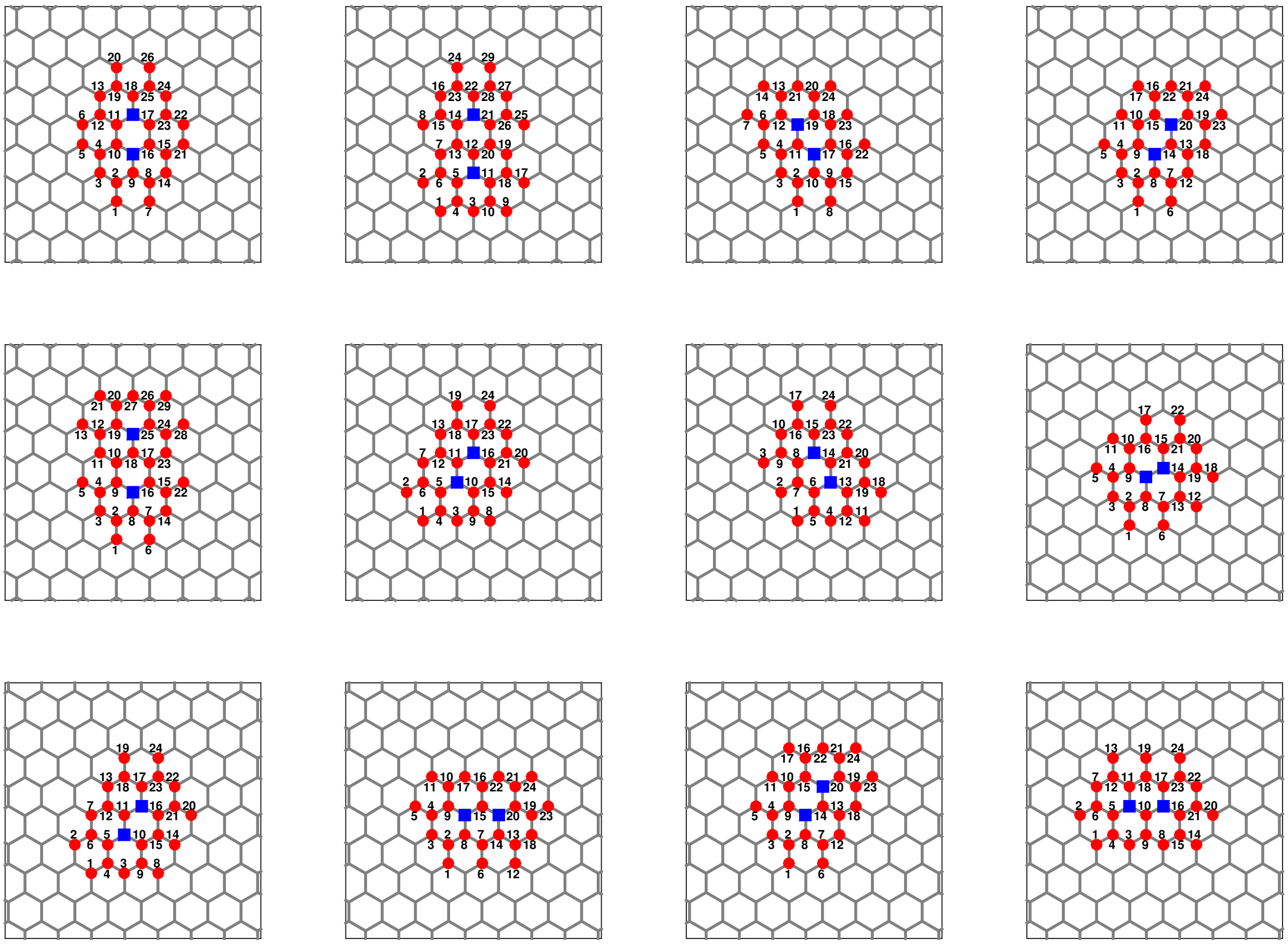

FIG. S3. 2 non-mutant and 10 mutant families of Trimer generation in flat graphene. 12 families consist of 298 candidates are generated from 2 parent.

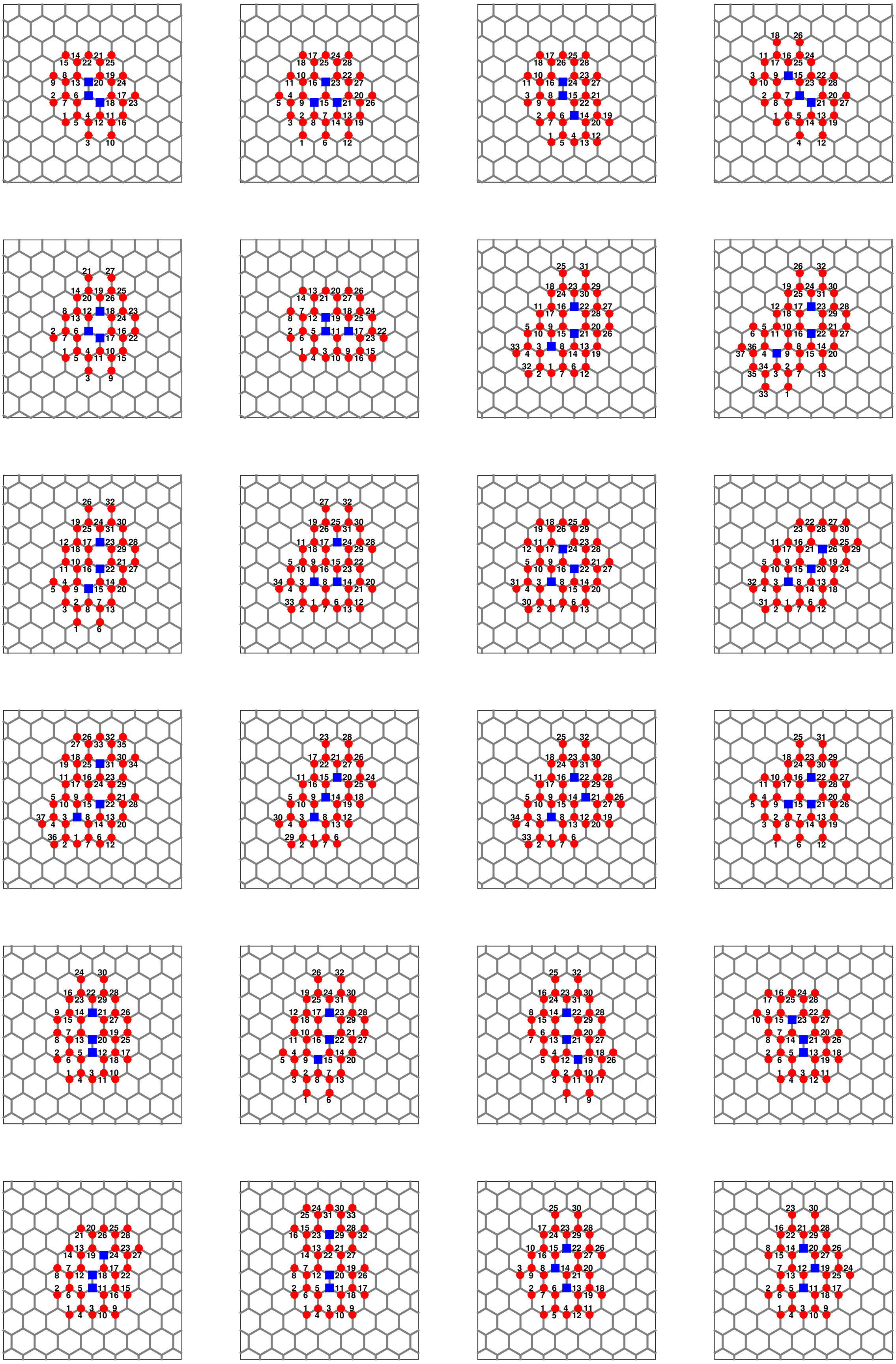

FIG. S4. 3 non-mutant and 21 mutant families of tetramer generation in flat graphene. 24 families consist of 737 candidates are generated from 3 parents.

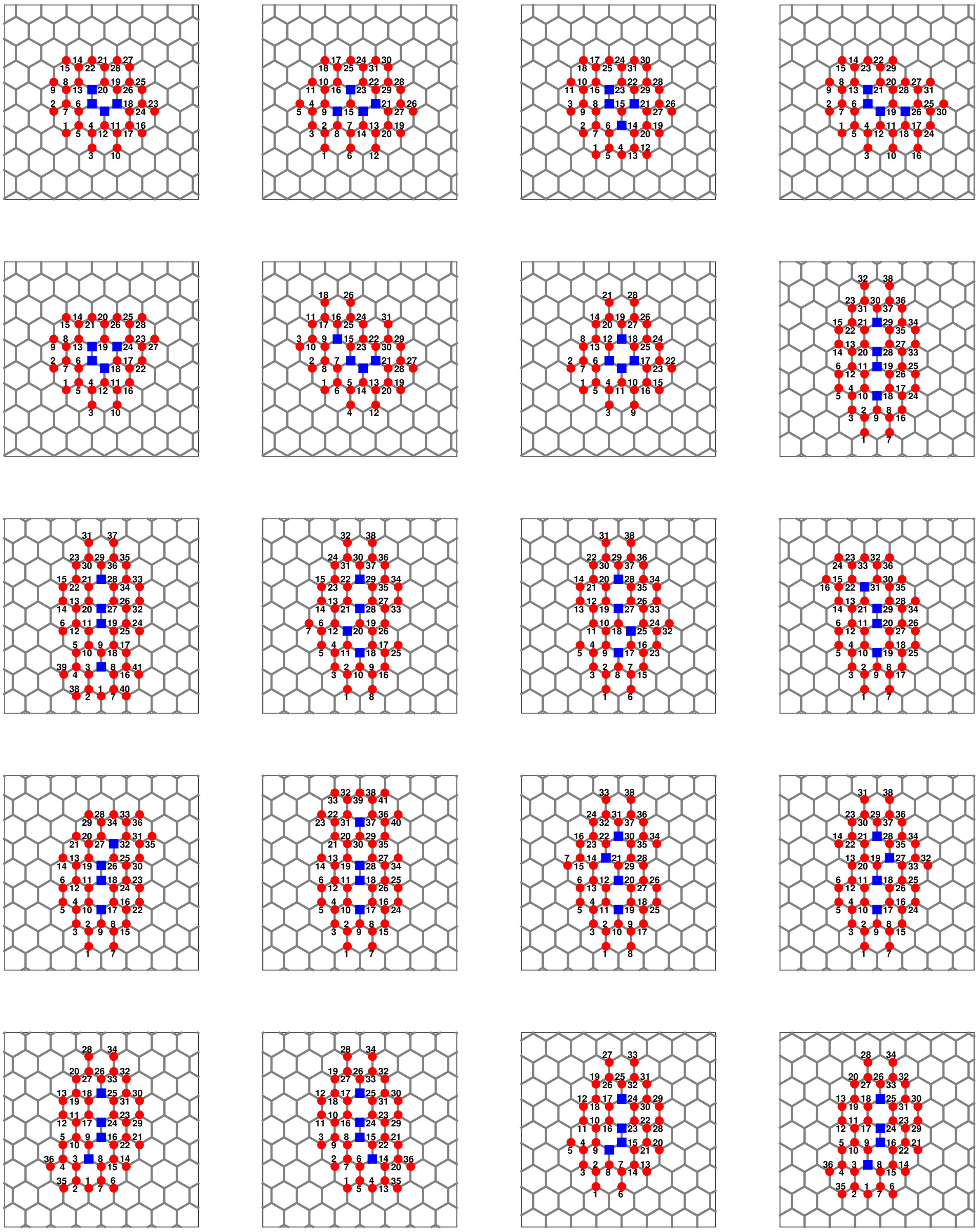

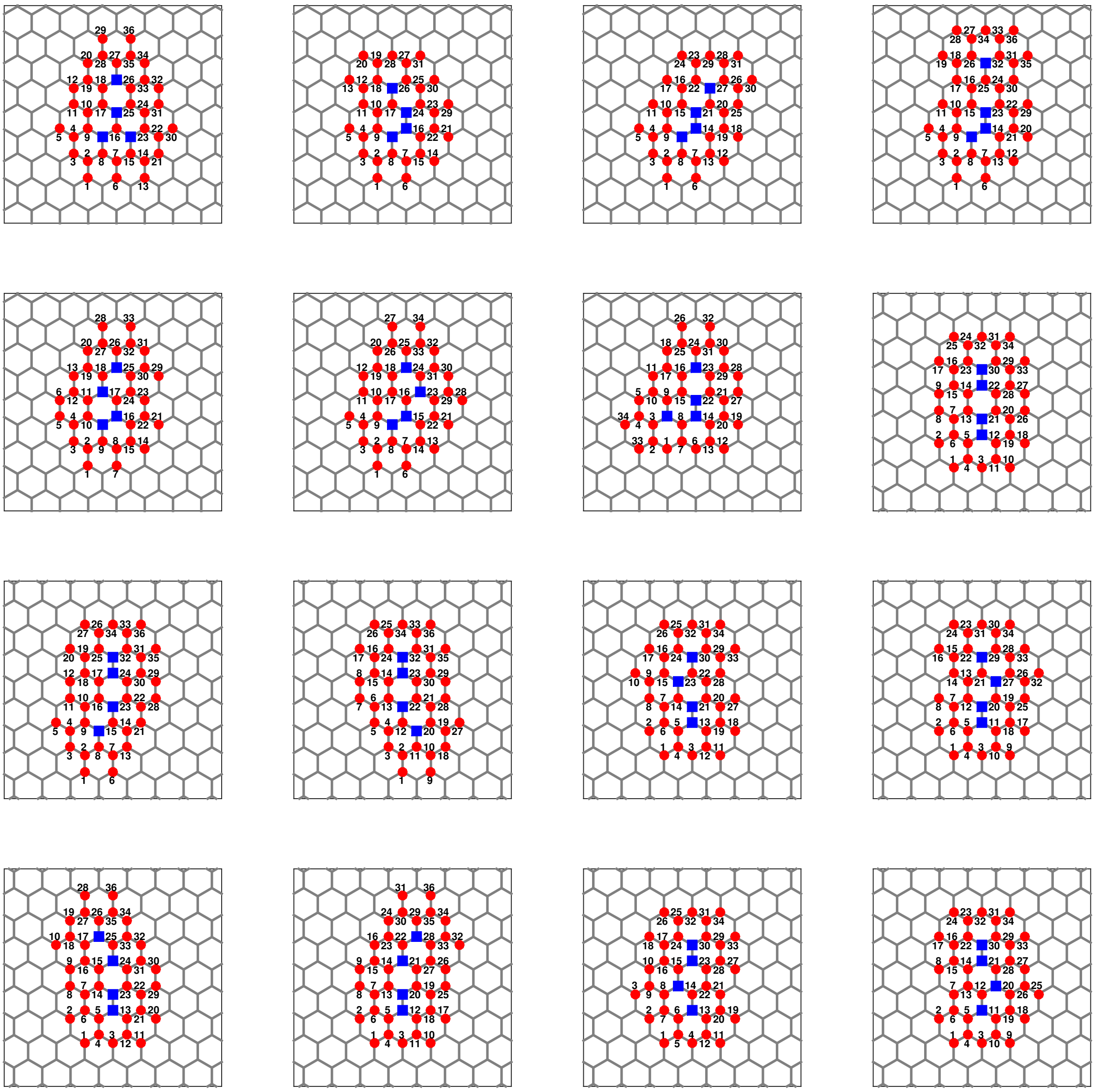

FIG. S5. 4 non-mutant and 32 mutant families of Pentamer generation in flat graphene. 36 families consist of 1242 candidates are generated from 4 parents.

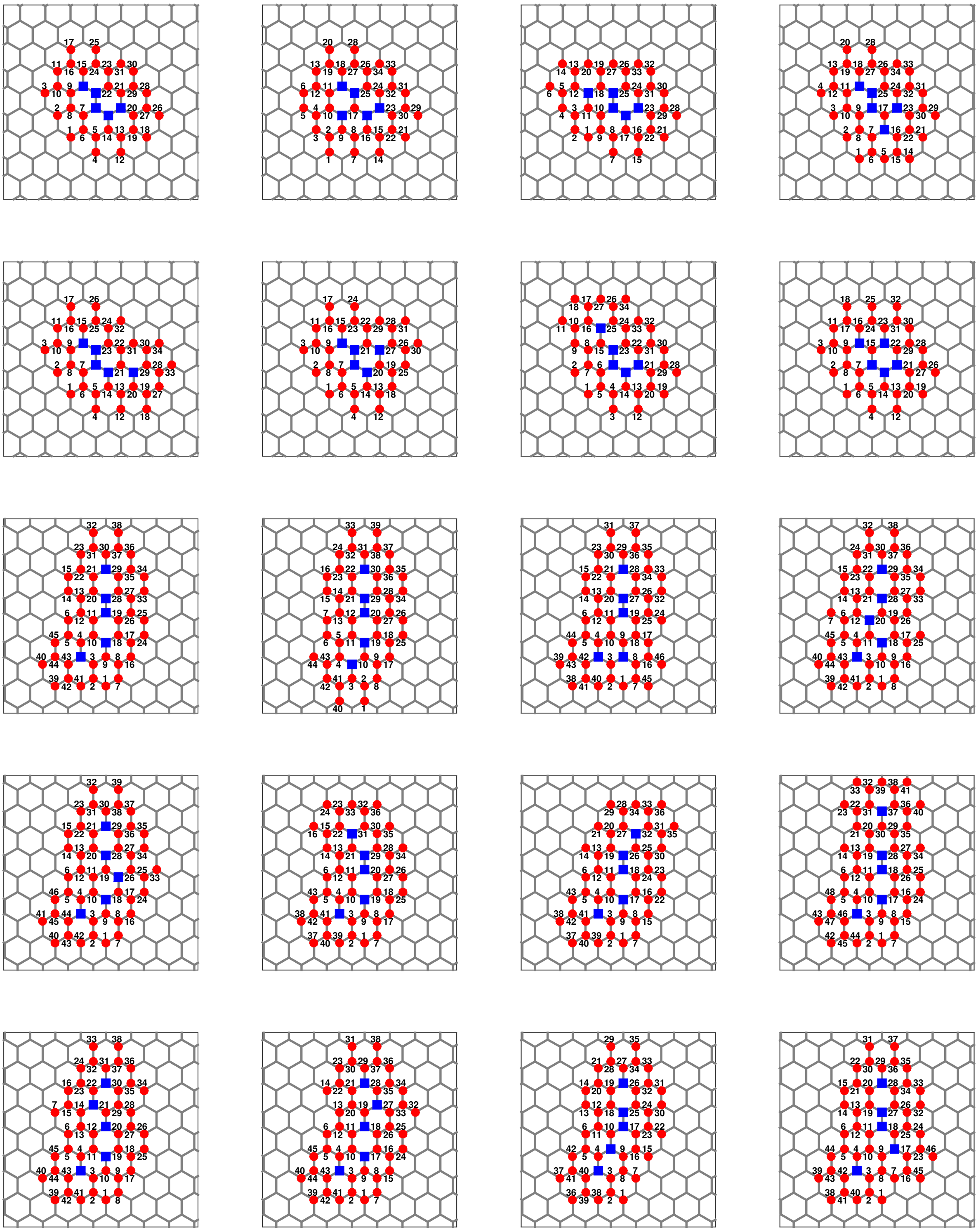

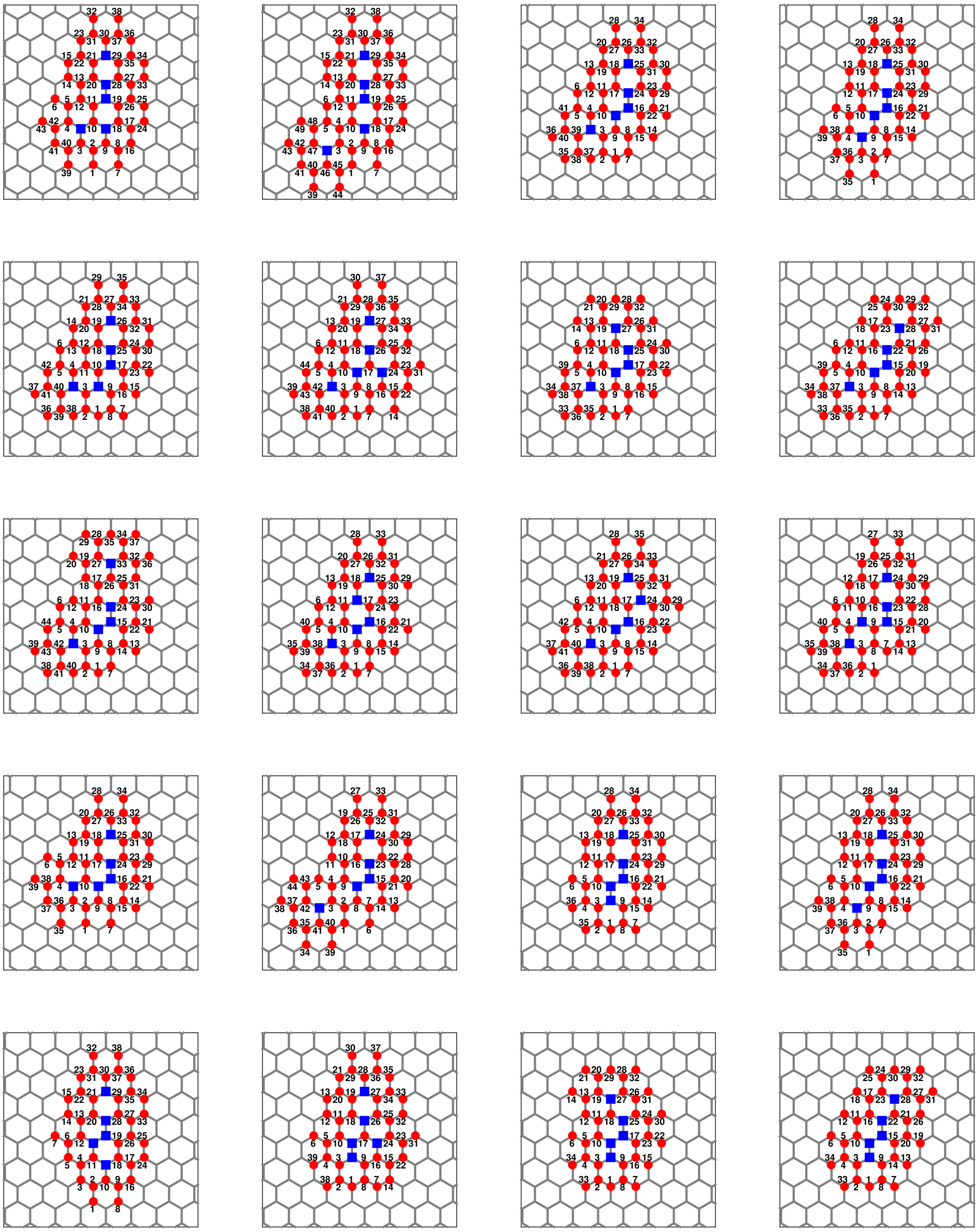

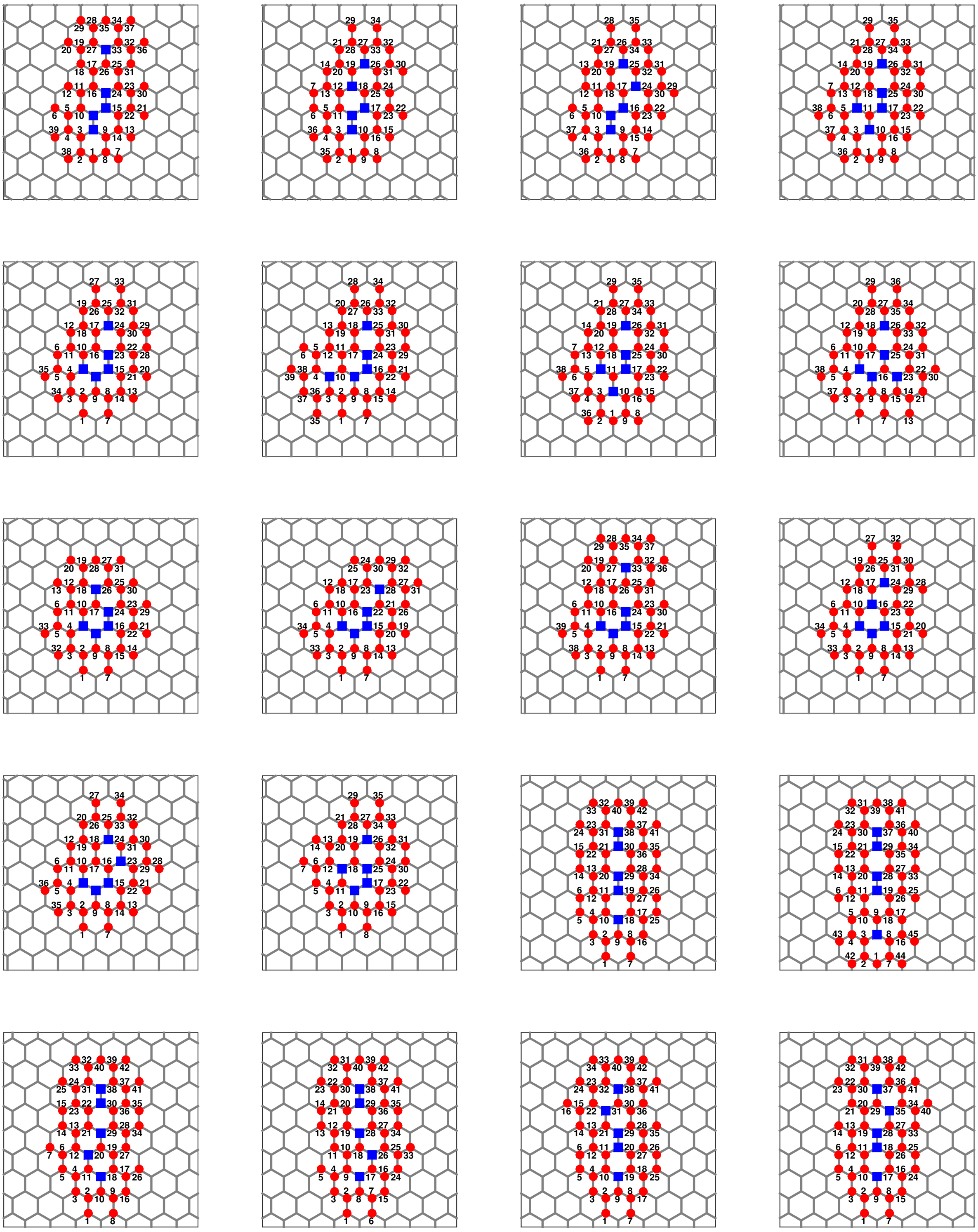

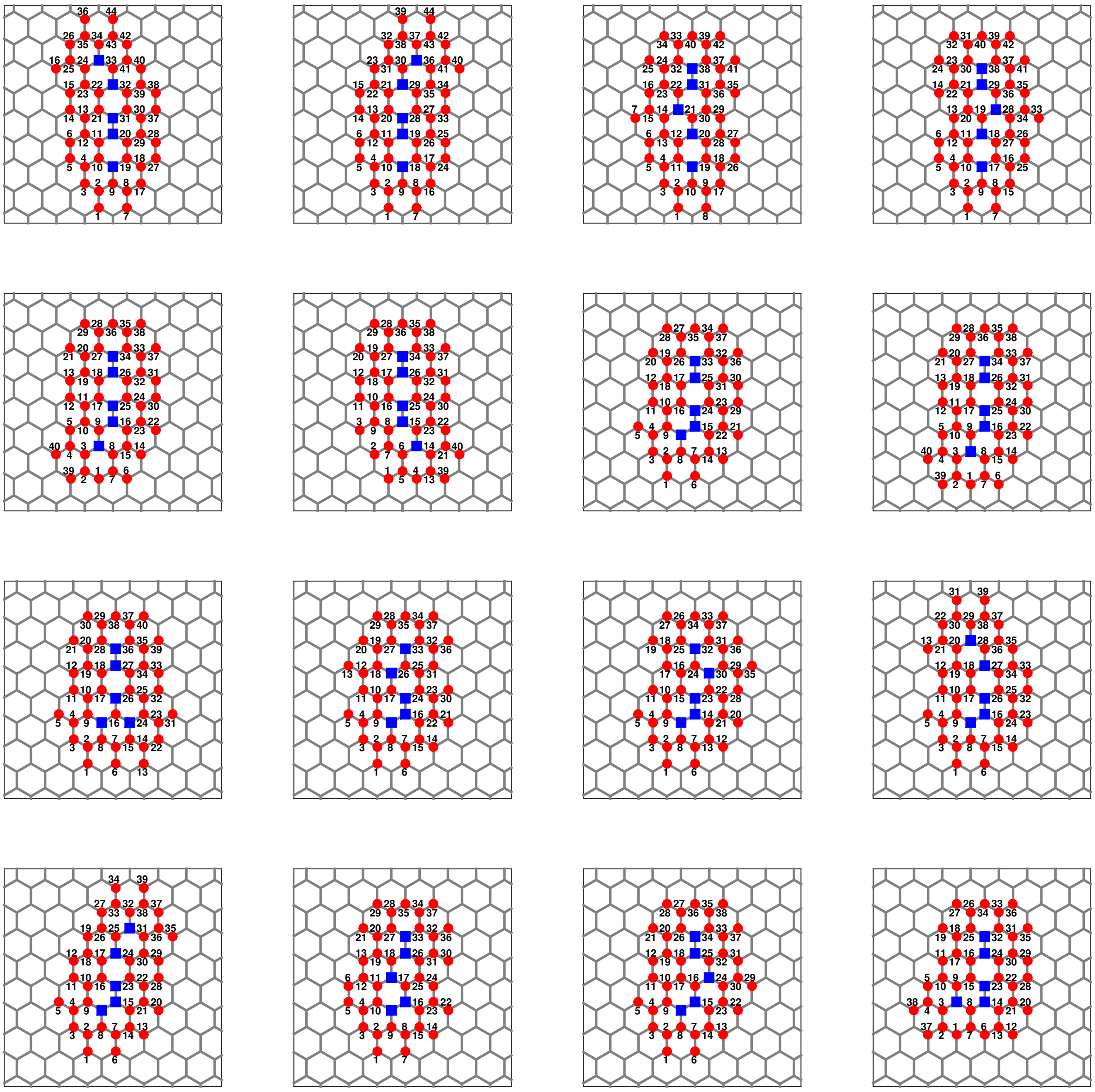

FIG. S6. 7 non-mutant and 69 mutant families of Hexamer generation in flat graphene. 76 families consist of 3006 candidates are generated from 7 parents.

TABLE S-I. Relative total energies (in eV) of configurations which are succeeded at the pre-selection stage on flat graphene. The configuration numbers correspond to those in Figure 4.

| Dimers | | | Trimers | | | Tetramers | | | Pentamers | | | Hexamers | | |
|---|---|---|---|---|---|---|---|---|---|---|---|---|---|---|
| Config | Parent | Relative Energy | Config | Parent | Relative Energy | Config | Parent | Relative Energy | Config | Parent | Relative Energy | Config | Parent | Relative Energy |
| 1 | D | 0.059 | 7 | $D_4Tr$ | 0.142 | 12 | $D_4Tr_7Te$ | 0.287 | 21 | $D_4Tr_7Te_{14}P$ | 0.064 | 32 | $D_4Tr_7Te_{14}P_{21}H$ | 0.096 |
| 2 | D | 0.000 | 8 | $D_4Tr$ | 0.000 | 13 | $D_4Tr_7Te$ | 0.240 | 22 | $D_4Tr_7Te_{14}P$ | 0.099 | 33 | $D_4Tr_8Te_{18}P_{26}H$ | 0.314 |
| 3 | D | 0.000 | 9 | $D_4Tr$ | 0.106 | 14 | $D_4Tr_7Te$ | 0.145 | 23 | $D_4Tr_7Te_{14}P$ | 0.168 | 34 | $D_4Tr_8Te_{18}P_{26}H$ | 0.295 |
| 4 | D | 0.059 | 10 | $D_2Tr$ | 0.000 | 15 | $D_4Tr_8Te$ | 0.240 | 24 | $D_4Tr_7Te_{14}P$ | 0.083 | 35 | $D_4Tr_8Te_{18}P_{26}H$ | 0.256 |
| 5 | D | 0.000 | 11 | $D_2Tr$ | 0.106 | 16 | $D_4Tr_8Te$ | 0.145 | 25 | $D_4Tr_7Te_{14}P$ | 0.204 | 36 | $D_4Tr_8Te_{18}P_{26}H$ | 0.140 |
| 6 | D | 0.059 | | | | 17 | $D_4Tr_8Te$ | 0.062 | 26 | $D_4Tr_8Te_{18}P$ | 0.167 | 37 | $D_4Tr_7Te_{14}P_{23}H$ | 0.161 |
| | | | | | | 18 | $D_4Tr_8Te$ | 0.113 | 27 | $D_4Tr_8Te_{18}P$ | 0.000 | 38 | $D_4Tr_7Te_{14}P_{23}H$ | 0.104 |
| | | | | | | 19 | $D_4Tr_9Te$ | 0.000 | 28 | $D_4Tr_8Te_{17}P$ | 0.083 | 39 | $D_4Tr_7Te_{14}P_{22}H$ | 0.227 |
| | | | | | | 20 | $D_4Tr_9Te$ | 0.145 | 29 | $D_4Tr_8Te_{17}P$ | 0.000 | 40 | $D_4Tr_7Te_{14}P_{22}H$ | 0.306 |
| | | | | | | | | | 30 | $D_4Tr_9Te_{19}P$ | 0.067 | 41 | $D_4Tr_7Te_{14}P_{22}H$ | 0.258 |
| | | | | | | | | | 31 | $D_4Tr_9Te_{19}P$ | 0.099 | 42 | $D_4Tr_7Te_{14}P_{22}H$ | 0.173 |
| | | | | | | | | | | | | 43 | $D_4Tr_8Te_{17}P_{29}H$ | 0.140 |
| | | | | | | | | | | | | 44 | $D_4Tr_8Te_{17}P_{29}H$ | 0.194 |
| | | | | | | | | | | | | 45 | $D_4Tr_8Te_{17}P_{29}H$ | 0.016 |
| | | | | | | | | | | | | 46 | $D_4Tr_8Te_{17}P_{28}H$ | 0.096 |
| | | | | | | | | | | | | 47 | $D_4Tr_8Te_{17}P_{28}H$ | 0.173 |
| | | | | | | | | | | | | 48 | $D_4Tr_8Te_{17}P_{28}H$ | 0.196 |
| | | | | | | | | | | | | 49 | $D_4Tr_8Te_{17}P_{28}H$ | 0.243 |
| | | | | | | | | | | | | 50 | $D_4Tr_8Te_{17}P_{28}H$ | 0.140 |
| | | | | | | | | | | | | 51 | $D_4Tr_8Te_{17}P_{28}H$ | 0.210 |
| | | | | | | | | | | | | 52 | $D_4Tr_9Te_{19}P_{30}H$ | 0.104 |
| | | | | | | | | | | | | 53 | $D_4Tr_9Te_{19}P_{30}H$ | 0.000 |
| | | | | | | | | | | | | 54 | $D_4Tr_9Te_{19}P_{30}H$ | 0.144 |

### B. High Throughput Calculations on Curved Graphene ($52^o$ and $90^o$)

**$52^o$-Bending.**

*Dimers*− For dimers, we consider 21 candidates, which are distributed evenly on the left and the right hand sides of the MRCL. Six configurations are found to be stable ($D_1$..$D_6$) at the pre-selection, two of them are eliminated during the pool selection. Since the left hand side (LHS) and right hand side (RHS) of MRCL are symmetric and also the parent of dimers is located on MRCL, $D_1$ is equivalent to $D_6$, and $D_2$ is equivalent to $D_5$. Therefore $D_1$ and $D_5$ are eliminated. Both $D_2$ and $D_4$ are in ortho-position but $D_4$ is positioned on MRCL while $D_2$ makes a $60^o$ angle with it. $D_3$ and $D_6$ are para-positioned. It should be noted that $D_3$ and $D_4$ will most likely generate the same configurations in trimers, which are located on MRCL. Nevertheless both $D_3$ and $D_4$ are investigated to ensure our consideration. Total energies and binding energies per H atom of 21 candidates are summarized in Table S-II and Table S-XII, respectively. The most favourable dimer configuration is found to be $D_4$ on curved graphene.

*Trimers*− 43 candidates are generated from 4 parents. Only 4 of them succeed after the selection process. 16 candidates are generated from $D_2$ (see $D_2$Tr in Figure S7), which are distributed non-symmetrically. $Tr_7$ and $Tr_{10}$, which form a line making $60^o$ angle with MRCL, are eliminated because the same configurations are obtained from $D_6Tr$. Both $Tr_8$ and $Tr_9$ are in O-O position, but with a small energy difference. This small energy difference resulting from the difference in curvature. This effect will play major roles in subsequent generations. Since both parents lie on the MRCL in $D_3$Tr, only 7 candidates are enough to investigate the stable configurations (see $D_3$Tr in Figure S7). In agreement with our prediction, both the P-O positioned $Tr_{11}$ and the O-P positioned $Tr_{13}$ give the same configuration, and therefore $Tr_{11}$ is eliminated. Similar to $D_3$Tr, 9 candidates are analyzed for $D_4$Tr (see $D_4$Tr in Figure S7). O-O positioned $Tr_{12}$ and $Tr_9$ are equivalent, hence $Tr_{12}$ is eliminated. $Tr_{13}$, which is in O-P position, contributes to the formation of a single line on MRCL. In family $D_6$Tr, there exist more candidates on the RHS compared to the LHS of MRCL, with 11 candidates (see $D_6$Tr in Figure S7). $Tr_{14}$ and $Tr_{15}$ are in P-O position and also they are aligned along single line making an angle $60^o$ with the MRCL. Binding energy of $Tr_{14}$ is higher than that of $Tr_{15}$ because $Tr_{14}$ is very close to the MRCL as seen in Table S-XIII. Both of them will give same configurations in next generation. For all trimer configurations considered, $Tr_{11}$ and $Tr_{14}$ are found to be the most stable configurations. Notice that, up to the third generation, single line formation on MRCL is favoured.

*Tetramers*− Tetramer generation consists of 94 candidates from 4 parents (see third row of Figure S7). From 22 candidates the only successful configurations are $Te_{16}$, $Te_{17}$ and $Te_{18}$ in $D_2Tr_8$Te. O-O-P positioned $Te_{16}$ tends to form a single line on $2^{nd}$ MRCL, which are the parallel lines on the left and on the right that consist of sites closest to the MRCL. O-O-O positioned $Te_{17}$ and $Te_{18}$ seem to be close to each other geometrically, but there is a considerable energy difference about 122 meV between them. $Te_{17}$ will most likely forms a double line (armchair geometry) on MRCL and its nearest neighbour. Additionally, there can be another possible configuration which is a closed hexagonal ring. $Te_{18}$ is less likely to form a single line where curvature changes because this geometry will be prevented by two parents on $2^{nd}$ MRCL. The other possible geometry can be a closed hexagonal like $Te_{17}$. In family $D_2Tr_9$Te, $Te_{19}$ and $Te_{21}$ have similar geometries with $Te_{18}$ and $Te_{16}$ respectively as well as having very close energies. For this reason both of them are eliminated. $Te_{20}$ is kept in order to observe the effect of relative position with respect to the MRCL in subsequent generation. 23 candidates are investigated for in $D_2Tr_{13}$Te family. $Te_{23}$ favours to form single line on MRCL and corresponds to O-P-O position. O-P-P positioned $Te_{22}$ deviates from MRCL to form a double line. O-P-P positioned $Te_{24}$ is eliminated because this configuration will reproduce the same candidates with $Te_{23}$, as it was the case for $D_3$ and $D_4$. 16 possible candidates have been explored for family $D_2Tr_8$Te and only one configuration is selected in P-O-O position, namely $Te_{27}$. It forms a short linear chain with a $60^o$-angle with the MRCL. Occupancies of A and B sublattices are balanced. $Te_{25}$ and $Te_{26}$, which correspond to P-O-P and P-O-O positions respectively, are unfavorable candidates because of the first criteria of pool-selection. It can be seen that from Table S-IV and Table S-XIV, $Te_{23}$ is found to be the most favorable configuration when tetramer generation is evaluated based on energetics of configurations.

*Pentamers*− In generating pentamers from tetramers, 108 candidates have been analyzed (see fourth and fifth rows of Figure S7). These candidates are derived from 7 parents, thus we examine 7 families in this generation. In $D_2Tr_8Te_{16}$P, 17 candidates are considered, $P_{28}$ and $P_{29}$ proceed to the next generation. $P_{30}$ fails due to the first criteria of pool-selection. $P_{28}$, which is in O-O-P-O position, forms a single on $2nd$ MRCL. $P_{29}$ which is also in O-O-P-O position represents deviation from the linear behaviour. In $D_2Tr_8Te_{17}$P, we considered 14 candidates but only the O-O-O-O positioned $P_{31}$ has been selected. It is clearly seen that armchair geometry is realized in $D_2Tr_8Te_{17}$P. $D_2Tr_8Te_{18}$P is an interesting family of configurations. Family coming from $Te_{18}$ becomes extinct in this step. $P_{32}$..$P_{36}$ are failed at the pool-selection stage. We should mention that elimination of $P_{32}$ and $P_{36}$ are due to curvature. In $D_2Tr_9Te_{20}$P, only $P_{37}$ succeeds among 16 candidates. There is a very small energy difference between $P_{31}$ and $P_{37}$. It is clearly seen that armchair geometry reappears in $P_{37}$. Among 19 candidates in $D_4Tr_{13}Te_{22}P$, 3 of them succeed to the next generation. $P_{39}$ and $P_{40}$ have similar geometries to $P_{29}$ and $P_{28}$, respectively. It is interesting to note that although $P_{30}$ is closer to the MRCL more than $P_{38}$, $P_{30}$ is eliminated. The main reason can be that most of the parents on MRCL lift carbon atoms and the formation of kink on MRCL makes it easier to adsorb hydrogen atom. The same configuration on flat graphene $P_{21}$ was also transferred to the next generation. Binding energy of $P_{38}$ is higher than $P_{30}$ can be seen in Table S-XV. There is an energy difference about 50 meV which seems reasonable according to the threshold value. $D_4Tr_{13}Te_{23}P$ family consists of 12 candidates. $P_{41}$ fails at the pool-selection stage. Hydrogen atoms do not deviate from MRCL and $P_{42}$ O-P-O-P forms a single line. $D_6Tr_{14}Te_{27}P$ is the second family in pentamers, which can not produce any successful candidates. $P_{43}$..$P_{48}$ are eliminated in the first criteria of pool-selection. In tetramers, it was possible to have linear configurations making an angle $60^o$ with the MRCL but in pentamers those formations are all eliminated and only lines along the MRCL are favored. This is a clear indication of the curvature effect. In $90^o$ bending, this effect becomes more pronounced, as it will be discussed below.

*Hexamers*− Finally, in generating hexamers from pentamers, 195 candidates from 9 parents are investigated (see sixth and seventh rows of Figure S7). Second criteria of the pool-selection which is related to the symmetry, was not taken into account in hexamers, because we want to see all possible preferential configurations. Family $D_2Tr_8Te_{16}P_{28}H$ consists of 24 candidates. Four of them are eliminated $H_{49}$, $H_{52}$, $H_{53}$, $H_{54}$ and two of them succeeded to be parents for the next generation. O-O-P-O-P positioned $H_{50}$ shows that, single line occurs preferentially at the $2nd$ MRCL. O-O-P-O-O positioned $H_{51}$ was also appeared in flat graphene as $H_{48}$ reveals that For $D_2Tr_8Te_{16}P_{29}H$, O-O-P-O-O positioned $H_{56}$ and O-O-P-O-P positioned $H_{57}$ are on the first and the second MRCL where in $H_{57}$ the number of hydrogens are equal on the first and the second MRCLs. In addition, there is a small but significant difference between these two successful candidates. Configurations in the $D_2Tr_8Te_{17}P_{31}H$ family are among the most stable ones. O-O-O-O-O positioned $H_{58}$ sets a perfect armchair pattern at the edge of the curved graphene. O-O-O-O-P positioned $H_{59}$ is quite similar to $H_{56}$ and it is possible to continue with single or double lines. Interestingly, O-O-O-O-O positioned $H_{60}$ is oriented to from a triple line on MRCL and nearest neighbours. Similarly, $D_2Tr_9Te_{20}P_{37}H$ generates stable armchair geometries. On the other hand, $D_4Tr_{13}Te_{22}P_{38}H$ family does not have any successful candidates. Although 4 parents which are located along lines making $60^o$ angle with the MRCL, their derivatives were stable in the tetramer generation, with two additional atoms on the MRCL can not succeed. This is because of the presence of single line and armchair configurations in the pool, which are very stable. $D_4Tr_{13}Te_{22}P_{39}H$ behaves very similar to $D_4Tr_{13}Te_{22}P_{29}H$ in terms of failed and succeeded candidates. Some candidates in $D_4Tr_{13}Te_{22}P_{40}H$ and $D_4Tr_{13}Te_{22}P_{28}H$ have similar geometries but there are differences between failed candidates. The main reason is $D_4Tr_{13}Te_{22}P_{28}H$ being on the second MRCL while $D_4Tr_{13}Te_{22}P_{40}H$ is on the first. In $D_4Tr_{13}Te_{23}P_{42}H$, $H_{74}$..$H_{76}$ are the successful candidates. O-P-O-P-O positioned $H_{74}$ marks a deviation from the linear succession. Still, linear hydrogen chain formation continues on MRCL with $H_{75}$, which is created by O-P-O-P-O positioning. Similar single line formation on MRCL created by O-P-O-P-P positioning with $H_{76}$ but with a slight difference that there are unpaired sublattice sites. Finally, we show $D_4Tr_{16}Te_{27}P_{43}H$, whose parent was eliminated at the pentamer generation. We continued to investigate this configuration to demonstrate that the configurations making an angle with the MRCL are not feasible, even if their counterparts on flat graphene are energetically most favorable ones. As a summary of this section, the first remarkable observation is that the most stable configurations have armchair pattern, namely $H_{58}$ and $H_{64}$. They both evolve from $D_2$ but following different lines starting from the trimer generation. In addition to armchair geometry, single line formation is also favored. Remarkably, single line formation is preferable when there are no unpaired hydrogens, whereas armchair orientation takes the minimum energy with unpaired hydrogens. Another important result is that there is a limit on the length of the single line when it makes $60^o$ angle with the MRCL. This is mainly due to the changing RoC in this direction. For the $52^o$ bending, maximum length of the single lines is found to be tetramers. In what follows, we investigate the most favourable hydrogen configurations $90^o$ bending.

**$90^o$-Bending**. All-tested configurations on $90^o$-bent graphene are presented in Figure S8) with those of $90^o$-bent graphene, eliminated and successful candidates during the selection process are shown in Figure 7. The first difference between $52^o$ and $90^o$ appears in the trimer generation. $Tr_{14}$ is suppressed due to the stronger curvature effect in $90^o$-bending, which succeeded in $52^o$. This means that only lines as short as a hydrogen dimer is possible if it is not aligned with the MRCL. Configuration $Te_{18}(52^o)$ is eliminated in $90^o$-bending in the pre-selection stage thus, this configuration is not marked with black triangle. Configuration $Te_{18}(90^o)$ is eliminated at the first criteria in the pool-selection. The only different candidate in $D_2Tr_9Te(90^o)$ family is $Te_{20}$. $Te_{18}(90^o)$ and $Te_{20}(90^o)$ have exactly the same geometries. The family $D_6Tr_{14}Te(90^o)$ was eliminated by the end of trimers generation. If it was transferred to the next generation, it would have the same candidates as in $52^o$-bending. In family $D_2Tr_8Te_{16}P$, there is no difference between successful candidates belonging to different bending angles. However, $P_{30}(52^o)$, which is eliminated at the second stage, does not appear in the evolution on $90^o$-bent graphene. Family $D_2Tr_8Te_{18}P(52^o)$ was not observed in $90^o$-bending, because its parent was eliminated before. Family $D_4Tr_{13}Te_{23}P(90^o)$ is similar to $D_4Tr_{13}Te_{22}P(52^o)$ except the configuration $P_{38}$, which is eliminated due to stronger curvature. The main differences between $52^o$ and $90^o$ bending become clear in hexamers. Even though $D_2Tr_8Te_{16}P_{28}H(52^o)$ and $D_2Tr_8Te_{16}P_{29}H(90^o)$ originate from the same parents, higher curvature does not allow those configurations which are not on the MRCL. The geometry constituted by both $H_{50}(52^o)$ and $H_{51}(52^o)$, does not appear in $90^o$-bending. In $52^o$-bending, hydrogens can prefer to be ordered along a short line which makes an angle $60^o$ with the MRCL within family while in $90^o$-bending there is no such geometry. $H_{52}(52^o)$ which was eliminated, appears as a successful candidate $H_{43}(90^o)$. This geometrical change reveals that linear chain formation on the $2^{nd}$ MRCL does not continue in $90^o$ bending. In the family $D_2Tr_8Te_{16}P_{30}H$, $H_{44}$ which is same with $H_{55}(52^o)$, creates a single line on $2^{nd}$ MRCL. The geometry which is created by $H_{57}(52^o)$ is not included in $90^o$ bending due to the elimination at the pre-selection stage. Families $D_2Tr_8Te_{17}P_{31}H$ and $D_2Tr_9Te_{21}P_{32}H$ have similar geometries. It should be also noted that $H_{48}$ which forms armchair geometry in $90^o$ bending, fails in $52^o$ bending because this configuration can not pass the pre-selection. $D_2Tr_8Te_{16}P_{29}H(90^o)$ and $D_4Tr_{13}Te_{23}P_{34}H(90^o)$ have similar geometries. Both $H_{58}$ and $H_{59}$ form a single line on MRCL in the $D_4Tr_{13}Te_{24}P_{36}H$ family. $H_{59}$ leaves an unoccupied site for this reason, $H_{59}$ is less favoured in total energy when compared to $H_{58}$. $H_{57}(90^o)$ is eliminated during the pool selection. However, $H_{74}(52^o)$ which is identical with $H_{57}(90^o)$ pass the whole selection process. This difference is resulting from the strength of a kink formation on MRCL in $90^o$-bending more than $52^o$-bending. The last family $D_6Tr_{14}Te_{28}P_{37}H(90^o)$ is similar with $D_6Tr_{14}Te_{27}P_{43}H(52^o)$. In summary, the first dramatic result is that increase in bending angle leads to the formation of short dimers on a single line which makes an angle $60^o$ with MRCL. Number of eliminated candidates, especially in hexamer generation, are less than those of $52^o$ bending due to the decrease in RoC. This result indicates that families of pentamer and hexamer generations are more stable than $52^o$ bending. Similar to $52^o$-bent graphene, armchair geometry is more favourable than linear chain formation along MRCL. Binding energy per H atom in armchair direction is about 40 meV more than that of linear direction.

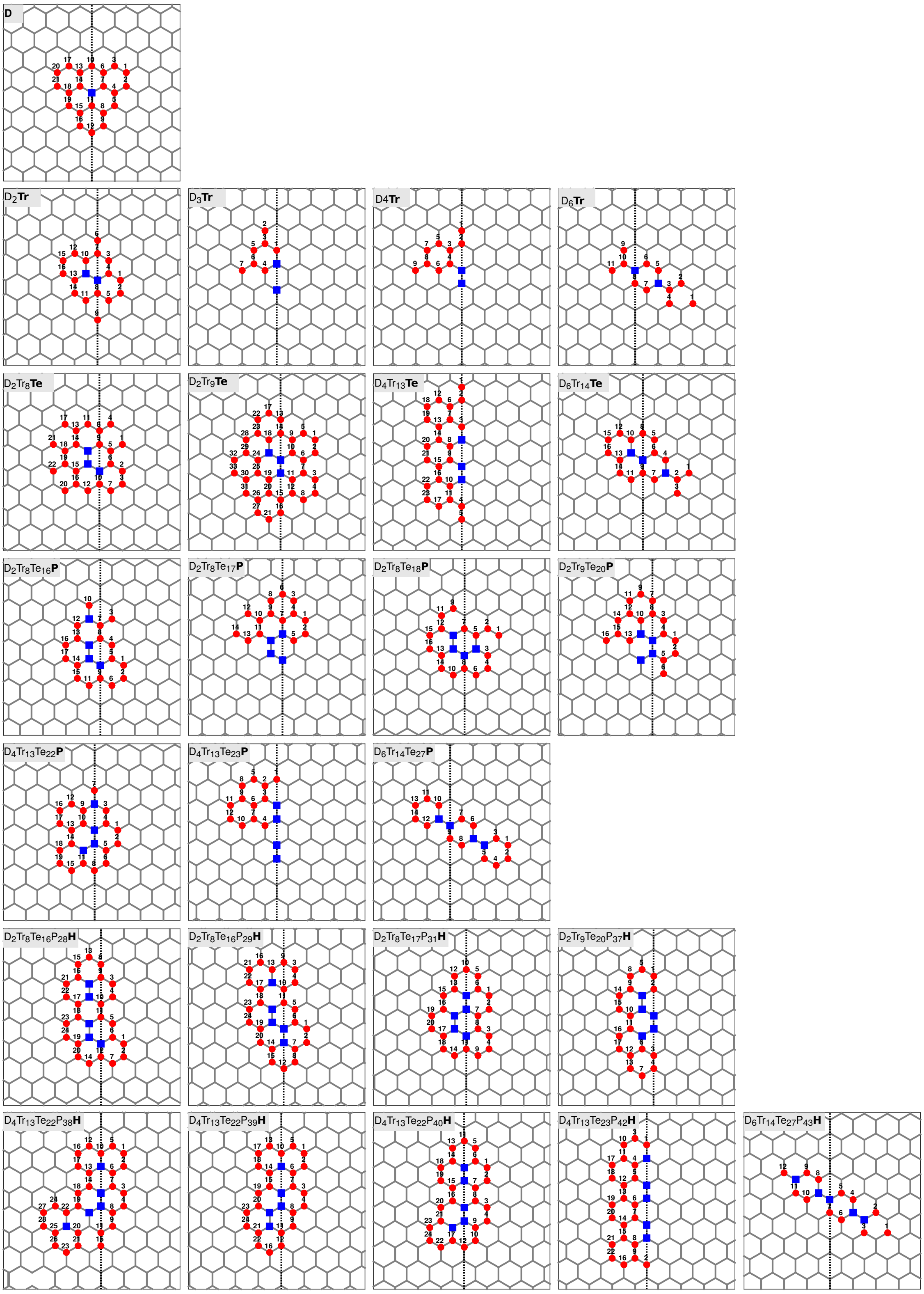

FIG. S7. Candidates on curved graphene with 52$^o$ bending.

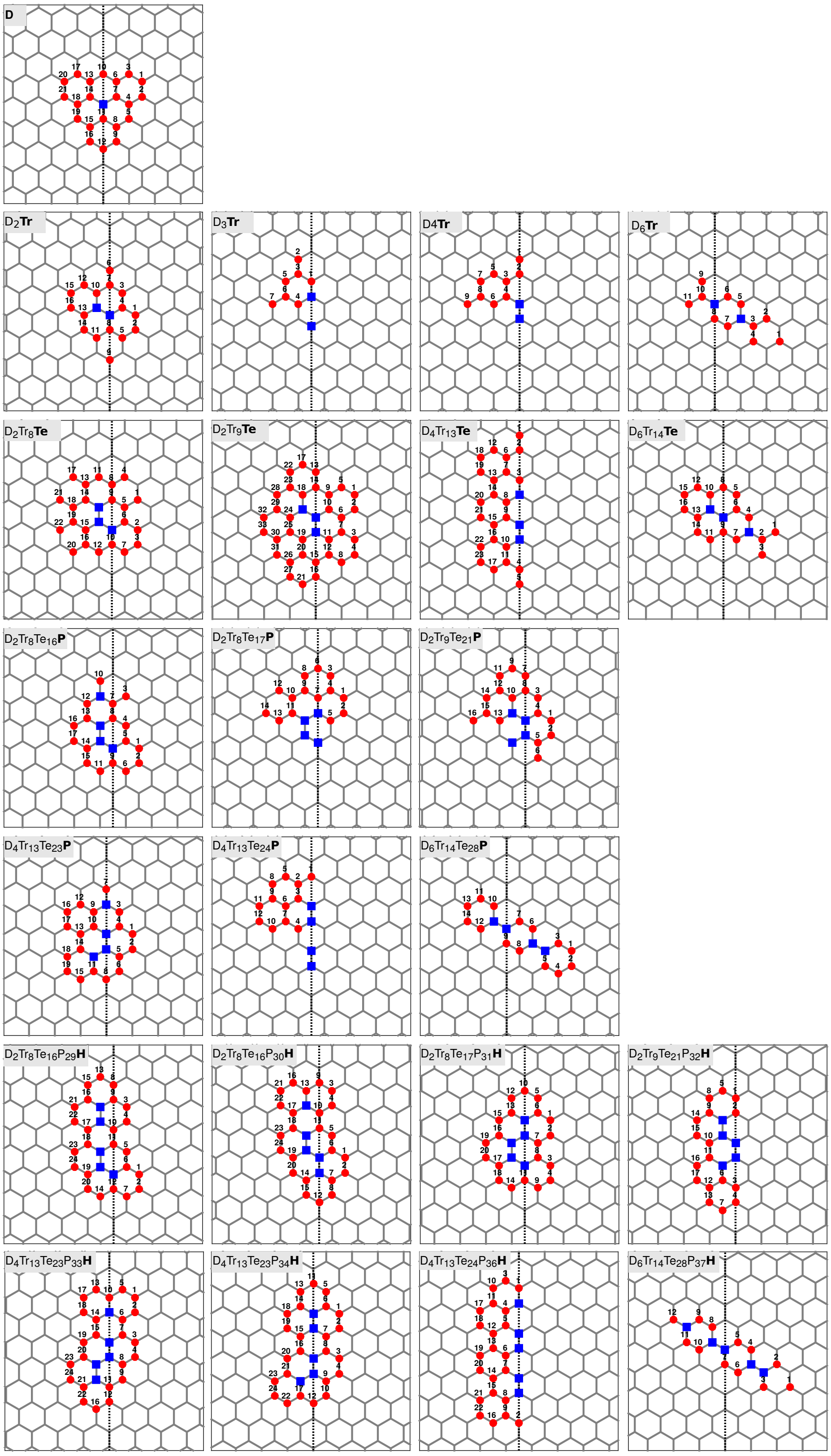

FIG. S8. Candidates on curved graphene with 90° bending.

TABLE S-II. Total energies of Dimer configurations on $52^o$-bent graphene with respect to the minimum energy configuration in eV. The configuration numbers correspond to those in Figure S7.

| Config | D |
|---|---|
| 1 | 1.345 |
| 2 | 1.010 |
| 3 | 1.019 |
| 4 | 1.266 |
| 5 | 0.121 |
| 6 | 1.205 |
| 7 | 0.058 |
| 8 | 1.204 |
| 9 | 0.998 |
| 10 | 0.092 |
| 11 | 0.000 |
| 12 | 1.307 |
| 13 | 1.201 |
| 14 | 0.064 |
| 15 | 1.206 |
| 16 | 0.995 |
| 17 | 1.013 |
| 18 | 1.263 |
| 19 | 0.121 |
| 20 | 1.338 |
| 21 | 1.003 |

TABLE S-III. Relative total energies (in eV) of 4 family which consists of 43 candidates in Trimer generation on $52^o$ bent system. The configuration numbers correspond to those in Figure S7.

| Config | $D_2$Tr | $D_3$Tr | $D_4$Tr | $D_6$Tr |
|---|---|---|---|---|
| 1 | 0.843 | 0.000 | 0.814 | 1.019 |
| 2 | 0.138 | 0.765 | 0.008 | 0.895 |
| 3 | 0.634 | 0.782 | 0.680 | 0.149 |
| 4 | 0.294 | 0.605 | 0.141 | 0.837 |
| 5 | 0.775 | 0.225 | 0.781 | 0.623 |
| 6 | 0.728 | 0.782 | 0.610 | 0.642 |
| 7 | 0.611 | 0.997 | 0.857 | 0.642 |
| 8 | 0.141 | | 0.675 | 0.612 |
| 9 | 0.888 | | 0.563 | 0.821 |
| 10 | 0.144 | | | 0.139 |
| 11 | 0.610 | | | 0.884 |
| 12 | 0.764 | | | |
| 13 | 0.293 | | | |
| 14 | 0.633 | | | |
| 15 | 0.134 | | | |
| 16 | 0.839 | | | |

TABLE S-IV. Relative total energies (in eV) of 4 family which consists of 94 candidates in Tetramer generation on 52$^o$ bent system. The configuration numbers correspond to those in Figure S7.

| Config | $D_2Tr_8Te$ | $D_2Tr_9Te$ | $D_4Tr_{13}Te$ | $D_6Tr_{14}Te$ |
|---|---|---|---|---|
| 1 | 1.260 | 1.449 | 0.820 | 1.371 |
| 2 | 1.454 | 1.424 | 1.278 | 0.153 |
| 3 | 0.237 | 1.059 | 0.000 | 1.292 |
| 4 | 0.944 | 1.439 | 0.078 | 0.757 |
| 5 | 1.602 | 1.435 | 1.296 | 0.711 |
| 6 | 0.144 | 0.896 | 0.965 | 1.332 |
| 7 | 1.265 | 1.213 | 1.125 | 0.748 |
| 8 | 1.336 | 1.298 | 0.600 | 1.230 |
| 9 | 0.022 | 0.843 | 1.105 | 1.213 |
| 10 | 0.234 | 1.056 | 0.154 | 0.239 |
| 11 | 0.148 | 0.315 | 1.109 | 0.675 |
| 12 | 1.109 | 1.228 | 1.135 | 1.577 |
| 13 | 1.226 | 0.984 | 0.264 | 0.425 |
| 14 | 0.308 | 1.105 | 1.277 | 1.265 |
| 15 | 1.057 | 0.152 | 0.610 | 0.316 |
| 16 | 0.837 | 1.389 | 1.157 | 1.378 |
| 17 | 1.303 | 1.380 | 1.112 | |
| 18 | 1.215 | 0.238 | 0.695 | |
| 19 | 0.890 | 0.028 | 1.346 | |
| 20 | 1.430 | 1.339 | 1.224 | |
| 21 | 1.042 | 0.696 | 1.471 | |
| 22 | 1.423 | 1.247 | 0.961 | |
| 23 | | 1.264 | 1.347 | |
| 24 | | 0.147 | | |
| 25 | | 1.603 | | |
| 26 | | 0.945 | | |
| 27 | | 1.278 | | |
| 28 | | 0.248 | | |
| 29 | | 1.467 | | |
| 30 | | 1.261 | | |
| 31 | | 1.485 | | |
| 32 | | 0.964 | | |
| 33 | | 1.512 | | |

TABLE S-V. Relative total energies (in eV) of 7 family which consists of 108 candidates in Pentamer generation on 52$^o$ bent system. The configuration numbers correspond to those in Figure S7

| Config | $D_2Tr_8Te_{16}P$ | $D_2Tr_8Te_{17}P$ | $D_2Tr_8Te_{18}P$ | $D_2Tr_9Te_{20}P$ | $D_4Tr_{13}Te_22P$ | $D_4Tr_{13}Te_{23}P$ | $D_6Tr_{14}Te_{27}P$ |
|---|---|---|---|---|---|---|---|
| 1 | 1.153 | 0.304 | 0.473 | 0.748 | 0.670 | 0 | 1.196 |
| 2 | 0.229 | 0.869 | 1.289 | 0.748 | 0.922 | 0.761 | 0.399 |
| 3 | 0.858 | 0.881 | 0.480 | 0.619 | 0.593 | 0.102 | 0.533 |
| 4 | 1.253 | 0.755 | 0.816 | 0.561 | 0.836 | 0.689 | 1.124 |
| 5 | 0.347 | 0.884 | 0.945 | 0.561 | 0.595 | 0.890 | 0.372 |
| 6 | 0.927 | 0.934 | 0.721 | 0.620 | 0.747 | 0.816 | 0.994 |
| 7 | 0.893 | 0.094 | 0.885 | 0.778 | 0.052 | 0.701 | 1.068 |
| 8 | 0.618 | 0.762 | 0.505 | 0.618 | 0.726 | 1.000 | 1.059 |
| 9 | 0.151 | 0.612 | 0.348 | 0.924 | 0.853 | 0.790 | 0.916 |
| 10 | 0.08 | 0.616 | 0.751 | 0.092 | 0.577 | 1.147 | 0.346 |
| 11 | 0.757 | 0.552 | 0.867 | 0.888 | 0.118 | 0.795 | 1.102 |
| 12 | 0.635 | 0.968 | 0.340 | 0.751 | 0.818 | 0.643 | 0.528 |
| 13 | 0.872 | 0.742 | 0.744 | 0.890 | 1.225 | | 0.382 |
| 14 | 0.627 | 0.978 | 0.798 | 0.317 | 0.304 | | 1.183 |
| 15 | 0.779 | | 0.876 | 0.878 | 0.892 | | |
| 16 | 0.714 | | 0.838 | 1.134 | 1.157 | | |
| 17 | 0.953 | | | | 1.062 | | |
| 18 | | | | | 1.124 | | |
| 19 | | | | | 0.179 | | |

TABLE S-VI. Relative total energies (in eV) of 9 family which consists of 195 candidates in Hexamer generation on 52$^{o}$ bent system. The configuration numbers correspond to those in Figure S7.

| Config | $D_2Tr_8Te_{16}P_{28}H$ | $D_2Tr_8Te_{16}P_{29}H$ | $D_2Tr_8Te_{17}P_{31}H$ | $D_2Tr_9$ $Te_{20}P_{37}H$ | $D_2Tr_{13}Te_{22}P_{38}H$ | $D_2Tr_{13}Te_{22}P_{39}H$ | $D_4Tr_{13}Te_{22}P_{40}H$ | $D_4Tr_{13}Te_{23}P_{42}H$ | $D_6Tr_{14}Te_{27}P_{43}H$ |
|---|---|---|---|---|---|---|---|---|---|
| 1 | 1.485 | 1.148 | 1.294 | 1.258 | 0.542 | 0.459 | 1.281 | 0.044 | 0.637 |
| 2 | 0.264 | 1.224 | 0.811 | 0.013 | 1.544 | 1.501 | 0.958 | 0.112 | 0.757 |
| 3 | 1.328 | 0.621 | 1.444 | 1.007 | 0.907 | 0.741 | 1.201 | 1.134 | 0.581 |
| 4 | 0.607 | 1.300 | 0.465 | 1.019 | 1.438 | 1.472 | 1.055 | 0.669 | 1.525 |
| 5 | 1.668 | 1.436 | 1.174 | 0.072 | 1.366 | 1.280 | 1.213 | 1.068 | 1.244 |
| 6 | 0.329 | 1.043 | 0.156 | 0.202 | 0.840 | 0.677 | 0.310 | 0.744 | 1.583 |
| 7 | 1.226 | 0.362 | 1.557 | 1.344 | 1.301 | 1.244 | 1.128 | 1.080 | 1.150 |
| 8 | 1.213 | 1.200 | 1.031 | 1.153 | 0.783 | 0.419 | 0.898 | 0.190 | 1.440 |
| 9 | 0.184 | 1.246 | 1.191 | 0.142 | 1.369 | 1.671 | 0.980 | 1.097 | 1.085 |
| 10 | 1.308 | 1.076 | 0.078 | 1.538 | 0.273 | 0.206 | 0.896 | 0.331 | 1.874 |
| 11 | 0.512 | 1.013 | 0.234 | 1.028 | 0.780 | 0.070 | 0.199 | 1.292 | 1.118 |
| 12 | 0.213 | 0.192 | 1.256 | 1.173 | 1.360 | 1.344 | 1.068 | 0.796 | 0.599 |
| 13 | 0.179 | 0.211 | 0.000 | 1.218 | 1.067 | 1.246 | 1.228 | 1.201 | |
| 14 | 1.048 | 0.063 | 1.015 | 1.289 | 1.279 | 1.076 | 0.187 | 0.787 | |
| 15 | 1.192 | 1.339 | 1.576 | 0.808 | 1.229 | 1.016 | 1.322 | 1.182 | |
| 16 | 0.285 | 1.275 | 0.425 | 1.434 | 0.606 | 0.200 | 0.530 | 1.178 | |
| 17 | 1.113 | 0.659 | 0.979 | 0.449 | 1.502 | 0.607 | 0.217 | 1.231 | |
| 18 | 0.876 | 1.233 | 0.738 | | 1.399 | 1.297 | 1.341 | 1.533 | |
| 19 | 0.958 | 0.418 | 1.350 | | 1.397 | 1.426 | 0.620 | 1.436 | |
| 20 | 0.877 | 1.674 | 0.860 | | 0.833 | 1.030 | 1.672 | 1.524 | |
| 21 | 1.274 | 0.437 | | | 1.458 | 0.351 | 0.338 | 0.970 | |
| 22 | 0.936 | 1.486 | | | 1.117 | 1.190 | 1.222 | 1.362 | |
| 23 | 1.183 | 0.731 | | | 0.519 | 1.153 | 1.486 | | |
| 24 | 1.049 | 1.467 | | | 1.482 | 1.230 | 0.273 | | |
| 25 | | | | | 0.435 | | | | |
| 26 | | | | | 1.383 | | | | |
| 27 | | | | | 0.674 | | | | |
| 28 | | | | | 1.477 | | | | |

TABLE S-VII. Relative total energies (in eV) of 1 family which consists of 21 candidates in Dimer generation on 90° bent system. The configuration numbers correspond to those in Figure S8.

| Config | D |
|---|---|
| 1 | 1.445 |
| 2 | 1.091 |
| 3 | 1.082 |
| 4 | 1.412 |
| 5 | 0.189 |
| 6 | 1.296 |
| 7 | 0.112 |
| 8 | 1.296 |
| 9 | 1.058 |
| 10 | 0.129 |
| 11 | 0.000 |
| 12 | 1.813 |
| 13 | 1.292 |
| 14 | 0.109 |
| 15 | 1.292 |
| 16 | 1.052 |
| 17 | 1.074 |
| 18 | 1.407 |
| 19 | 0.181 |
| 20 | 1.435 |
| 21 | 1.083 |

TABLE S-VIII. Relative total energies (in eV) of 4 family which consists of 43 candidates in Trimer generation on 90° bent system. The configuration numbers correspond to those in Figure S8.

| Config | $D_2$Tr | $D_3$Tr | $D_4$Tr | $D_6$Tr |
|---|---|---|---|---|
| 1 | 0.971 | 0.000 | 0.805 | 1.130 |
| 2 | 0.240 | 0.802 | 0.005 | 1.046 |
| 3 | 0.714 | 0.832 | 0.677 | 0.260 |
| 4 | 0.422 | 0.668 | 0.150 | 0.931 |
| 5 | 0.839 | 0.298 | 0.808 | 0.709 |
| 6 | 0.762 | 0.869 | 0.685 | 0.733 |
| 7 | 0.674 | 1.053 | 0.935 | 0.732 |
| 8 | 0.150 | | 0.691 | 0.691 |
| 9 | 0.937 | | 0.609 | 0.903 |
| 10 | 0.149 | | | 0.240 |
| 11 | 0.669 | | | 1.022 |
| 12 | 0.836 | | | |
| 13 | 0.421 | | | |
| 14 | 0.716 | | | |
| 15 | 0.247 | | | |
| 16 | 0.969 | | | |

TABLE S-IX. Relative total energies (in eV) of 4 family which consists of 94 candidates in Tetramer generation on 90° bent system. The configuration numbers correspond to those in Figure S8.

| Config | $D_2Tr_8Te$ | $D_2Tr_9Te$ | $D_4Tr_{13}Te$ | $D_6Tr_{14}Te$ |
|---|---|---|---|---|
| 1 | 1.377 | 1.569 | 0.867 | 1.653 |
| 2 | 1.635 | 1.552 | 1.348 | 0.498 |
| 3 | 0.354 | 1.136 | 0.000 | 1.526 |
| 4 | 1.019 | 1.576 | 0.141 | 0.935 |
| 5 | 1.756 | 1.567 | 1.346 | 0.878 |
| 6 | 0.282 | 1.033 | 1.046 | 1.596 |
| 7 | 1.390 | 1.407 | 1.233 | 0.909 |
| 8 | 1.410 | 1.432 | 0.648 | 1.440 |
| 9 | 0.072 | 0.926 | 1.212 | 1.657 |
| 10 | 0.245 | 1.194 | 0.214 | 0.353 |
| 11 | 0.217 | 0.388 | 1.193 | 0.829 |
| 12 | 1.214 | 1.318 | 1.256 | 1.521 |
| 13 | 1.324 | 1.040 | 0.345 | 0.662 |
| 14 | 0.379 | 1.205 | 1.422 | 1.501 |
| 15 | 1.194 | 0.208 | 0.711 | 0.526 |
| 16 | 0.923 | 1.474 | 1.318 | 1.657 |
| 17 | 1.432 | 1.480 | 1.199 | |
| 18 | 1.406 | 0.258 | 0.782 | |
| 19 | 1.026 | 0.080 | 1.453 | |
| 20 | 1.572 | 1.407 | 1.318 | |
| 21 | 1.149 | 0.776 | 1.586 | |
| 22 | 1.561 | 1.327 | 1.065 | |
| 23 | | 1.392 | 1.443 | |
| 24 | | 0.282 | | |
| 25 | | 1.752 | | |
| 26 | | 1.021 | | |
| 27 | | 1.390 | | |
| 28 | | 0.357 | | |
| 29 | | 1.638 | | |
| 30 | | 1.388 | | |
| 31 | | 1.601 | | |
| 32 | | 1.059 | | |
| 33 | | 1.653 | | |

TABLE S-X. Relative total energies (in eV) of 6 family which consists of 92 candidates in Pentamer generation on 90° bent system. The configuration numbers correspond to those in Figure S8.

| Config | $D_2Tr_8Te_{16}P$ | $D_2Tr_8Te_{17}P$ | $D_2Tr_9Te_{21}P$ | $D_4Tr_{13}Te_{23}P$ | $D_4Tr_{13}Te_{24}P$ | $D_6Tr_{14}Te_{28}P$ |
|---|---|---|---|---|---|---|
| 1 | 1.295 | 0.396 | 0.883 | 0.836 | 0.000 | 1.500 |
| 2 | 0.333 | 0.991 | 0.883 | 1.083 | 0.746 | 0.668 |
| 3 | 0.958 | 0.910 | 0.676 | 0.682 | 0.110 | 0.833 |
| 4 | 1.359 | 0.815 | 0.663 | 0.942 | 0.677 | 1.364 |
| 5 | 0.476 | 0.987 | 0.653 | 0.722 | 0.936 | 0.553 |
| 6 | 1.006 | 0.964 | 0.675 | 0.842 | 0.874 | 1.207 |
| 7 | 0.920 | 0.073 | 0.801 | 0.103 | 0.758 | 1.318 |
| 8 | 0.667 | 0.787 | 0.672 | 0.822 | 1.010 | 1.175 |
| 9 | 0.157 | 0.656 | 0.968 | 0.930 | 0.799 | 1.183 |
| 10 | 0.090 | 0.662 | 0.075 | 0.667 | 1.166 | 0.519 |
| 11 | 0.810 | 0.634 | 0.919 | 0.167 | 0.834 | 1.327 |
| 12 | 0.670 | 1.054 | 0.806 | 0.959 | 0.674 | 0.813 |
| 13 | 0.952 | 0.874 | 0.987 | 1.361 | | 0.635 |
| 14 | 0.737 | 1.067 | 0.395 | 0.475 | | 1.473 |
| 15 | 0.854 | | 1.000 | 1.019 | | |
| 16 | 0.841 | | 1.229 | 1.263 | | |
| 17 | 1.087 | | | 1.191 | | |
| 18 | | | | 1.293 | | |
| 19 | | | | 0.324 | | |

TABLE S-XI. Relative total energies (in eV) of 8 family which consists of 167 candidates in Hexamer generation on 90° bent system. The configuration numbers correspond to those in Figure S8.

| Config | $D_2Tr_8Te_{16}P_{29}H$ | $D_2Tr_8Te_{16}P_{30}H$ | $D_2Tr_8Te_{17}P_{31}H$ | $D_2Tr_9Te_{21}P_{32}H$ | $D_4Tr_{13}Te_{23}P_{33}H$ | $D_4Tr_{13}Te_{23}P_{34}H$ | $D_4Tr_{13}Te_{24}P_{36}H$ | $D_6Tr_{14}Te_{28}P_{37}H$ |
|---|---|---|---|---|---|---|---|---|
| 1 | 1.612 | 1.253 | 1.411 | 1.264 | 0.518 | 1.417 | 0.022 | 1.004 |
| 2 | 0.356 | 1.384 | 0.879 | 0.005 | 1.628 | 1.047 | 0.151 | 1.155 |
| 3 | 1.449 | 0.676 | 1.566 | 1.067 | 0.838 | 1.339 | 1.208 | 0.861 |
| 4 | 0.685 | 1.411 | 0.500 | 1.017 | 1.607 | 1.153 | 0.689 | 1.869 |
| 5 | 1.769 | 1.496 | 1.193 | 0.082 | 1.378 | 1.259 | 1.134 | 1.606 |
| 6 | 0.418 | 1.134 | 0.182 | 0.148 | 0.693 | 0.356 | 0.736 | 1.957 |
| 7 | 1.321 | 0.420 | 1.610 | 1.389 | 1.336 | 1.186 | 1.132 | 1.421 |
| 8 | 1.257 | 1.258 | 1.065 | 1.177 | 0.483 | 0.917 | 0.233 | 1.766 |
| 9 | 0.211 | 1.328 | 1.253 | 0.167 | 1.780 | 1.067 | 1.149 | 1.394 |
| 10 | 1.349 | 1.094 | 0.087 | 1.596 | 0.205 | 0.939 | 0.877 | 1.988 |
| 11 | 0.506 | 1.068 | 0.155 | 1.053 | 0.094 | 0.231 | 1.404 | 1.444 |
| 12 | 0.194 | 0.229 | 1.260 | 1.248 | 1.369 | 1.132 | 0.854 | 0.954 |
| 13 | 0.218 | 0.200 | 0.000 | 1.231 | 1.332 | 1.261 | 1.312 | |
| 14 | 1.127 | 0.083 | 1.071 | 1.400 | 1.095 | 0.233 | 0.851 | |
| 15 | 1.234 | 1.356 | 1.659 | 0.881 | 1.071 | 1.357 | 1.308 | |
| 16 | 0.331 | 1.372 | 0.461 | 1.553 | 0.241 | 0.519 | 1.228 | |
| 17 | 1.164 | 0.684 | 1.061 | 0.496 | 0.675 | 0.199 | 1.290 | |
| 18 | 0.908 | 1.328 | 0.734 | | 1.410 | 1.463 | 1.598 | |
| 19 | 1.044 | 0.474 | 1.448 | | 1.492 | 0.692 | 1.499 | |
| 20 | 0.927 | 1.633 | 0.936 | | 1.124 | 1.771 | 1.576 | |
| 21 | 1.398 | 0.506 | | | 0.408 | 0.420 | 1.040 | |
| 22 | 1.029 | 1.613 | | | 1.251 | 1.327 | 1.436 | |
| 23 | 1.322 | 0.823 | | | 1.252 | 1.616 | | |
| 24 | 1.136 | 1.592 | | | 1.384 | 0.365 | | |

TABLE S-XII. Binding energies (in eV) per H atom of 1 family which consists of 21 candidates in Dimer generation on 52$^o$ bent system. The configuration numbers correspond to those in Figure S7.

| Config | D |
|---|---|
| 1 | 1.073 |
| 2 | 1.241 |
| 3 | 1.236 |
| 4 | 1.113 |
| 5 | 1.685 |
| 6 | 1.143 |
| 7 | 1.717 |
| 8 | 1.144 |
| 9 | 1.247 |
| 10 | 1.700 |
| 11 | 1.746 |
| 12 | 1.092 |
| 13 | 1.145 |
| 14 | 1.714 |
| 15 | 1.143 |
| 16 | 1.248 |
| 17 | 1.239 |
| 18 | 1.114 |
| 19 | 1.686 |
| 20 | 1.077 |
| 21 | 1.245 |

TABLE S-XIII. Binding energies (in eV) per H atom of 4 family which consists of 43 candidates in Trimer generation on 52$^o$ bent system. The configuration numbers correspond to those in Figure S7.

| Config | $D_2Tr$ | $D_3Tr$ | $D_4Tr$ | $D_6Tr$ |
|---|---|---|---|---|
| 1 | 1.503 | 1.784 | 1.512 | 1.444 |
| 2 | 1.738 | 1.529 | 1.781 | 1.486 |
| 3 | 1.572 | 1.523 | 1.557 | 1.734 |
| 4 | 1.686 | 1.582 | 1.737 | 1.505 |
| 5 | 1.525 | 1.709 | 1.523 | 1.576 |
| 6 | 1.541 | 1.523 | 1.581 | 1.570 |
| 7 | 1.580 | 1.451 | 1.498 | 1.570 |
| 8 | 1.737 | | 1.559 | 1.580 |
| 9 | 1.488 | | 1.596 | 1.510 |
| 10 | 1.736 | | | 1.737 |
| 11 | 1.580 | | | 1.489 |
| 12 | 1.529 | | | |
| 13 | 1.686 | | | |
| 14 | 1.573 | | | |
| 15 | 1.739 | | | |
| 16 | 1.504 | | | |

TABLE S-XIV. Binding energies (in eV) per H atom of 4 family which consists of 94 candidates in Tetramer generation on 52$^o$ bent system. The configuration numbers correspond to those in Figure S7.

| Config | $D_2Tr_8Te$ | $D_2Tr_9Te$ | $D_4Tr_{13}Te$ | $D_6Tr_{14}Te$ |
|---|---|---|---|---|
| 1 | 1.597 | 1.549 | 1.707 | 1.569 |
| 2 | 1.548 | 1.556 | 1.592 | 1.873 |
| 3 | 1.852 | 1.647 | 1.911 | 1.589 |
| 4 | 1.675 | 1.552 | 1.892 | 1.722 |
| 5 | 1.511 | 1.553 | 1.588 | 1.734 |
| 6 | 1.876 | 1.687 | 1.670 | 1.724 |
| 7 | 1.595 | 1.608 | 1.630 | 1.725 |
| 8 | 1.578 | 1.587 | 1.761 | 1.604 |
| 9 | 1.906 | 1.701 | 1.635 | 1.608 |
| 10 | 1.853 | 1.647 | 1.873 | 1.852 |
| 11 | 1.874 | 1.833 | 1.634 | 1.743 |
| 12 | 1.634 | 1.605 | 1.628 | 1.517 |
| 13 | 1.605 | 1.665 | 1.846 | 1.805 |
| 14 | 1.835 | 1.635 | 1.592 | 1.595 |
| 15 | 1.647 | 1.873 | 1.759 | 1.832 |
| 16 | 1.702 | 1.564 | 1.622 | 1.567 |
| 17 | 1.586 | 1.567 | 1.634 | |
| 18 | 1.608 | 1.852 | 1.738 | |
| 19 | 1.689 | 1.905 | 1.575 | |
| 20 | 1.554 | 1.577 | 1.606 | |
| 21 | 1.651 | 1.737 | 1.544 | |
| 22 | 1.556 | 1.600 | 1.671 | |
| 23 | | 1.596 | 1.575 | |
| 24 | | 1.875 | | |
| 25 | | 1.511 | | |
| 26 | | 1.675 | | |
| 27 | | 1.592 | | |
| 28 | | 1.850 | | |
| 29 | | 1.545 | | |
| 30 | | 1.596 | | |
| 31 | | 1.540 | | |
| 32 | | 1.670 | | |
| 33 | | 1.533 | | |

TABLE S-XV. Binding energies (in eV) per H atom of 7 family which consists of 108 candidates in Pentamer generation on 52$^o$ bent system. The configuration numbers correspond to those in Figure S7.

| Config | $D_2Tr_8Te_{16}P$ | $D_2Tr_8Te_{17}P$ | $D_2Tr_8Te_{18}P$ | $D_2Tr_9Te_{20}P$ | $D_4Tr_{13}Te_2 2P$ | $D_4Tr_{13}Te_{23}P$ | $D_6Tr_{14}Te_{27}P$ |
|---|---|---|---|---|---|---|---|
| 1 | 1.694 | 1.864 | 1.830 | 1.775 | 1.791 | 1.924 | 1.685 |
| 2 | 1.879 | 1.751 | 1.667 | 1.775 | 1.740 | 1.772 | 1.845 |
| 3 | 1.753 | 1.748 | 1.829 | 1.801 | 1.806 | 1.904 | 1.818 |
| 4 | 1.674 | 1.773 | 1.761 | 1.812 | 1.757 | 1.787 | 1.700 |
| 5 | 1.855 | 1.748 | 1.736 | 1.812 | 1.805 | 1.746 | 1.850 |
| 6 | 1.739 | 1.738 | 1.780 | 1.801 | 1.775 | 1.761 | 1.726 |
| 7 | 1.746 | 1.906 | 1.748 | 1.769 | 1.914 | 1.784 | 1.711 |
| 8 | 1.801 | 1.772 | 1.824 | 1.801 | 1.779 | 1.725 | 1.713 |
| 9 | 1.894 | 1.802 | 1.855 | 1.740 | 1.754 | 1.766 | 1.741 |
| 10 | 1.908 | 1.801 | 1.774 | 1.906 | 1.809 | 1.695 | 1.855 |
| 11 | 1.773 | 1.814 | 1.751 | 1.747 | 1.901 | 1.766 | 1.704 |
| 12 | 1.797 | 1.731 | 1.857 | 1.774 | 1.761 | 1.796 | 1.819 |
| 13 | 1.750 | 1.776 | 1.776 | 1.747 | 1.680 | | 1.848 |
| 14 | 1.799 | 1.729 | 1.765 | 1.861 | 1.864 | | 1.688 |
| 15 | 1.769 | | 1.749 | 1.749 | 1.746 | | |
| 16 | 1.782 | | 1.757 | 1.698 | 1.693 | | |
| 17 | 1.734 | | | | 1.712 | | |
| 18 | | | | | 1.700 | | |
| 19 | | | | | 1.889 | | |

TABLE S-XVI. Binding energies (in eV) per H atom of 9 family which consists of 195 candidates in Hexamer generation on 52$^o$ bent system. The configuration numbers correspond to those in Figure S7.

| Config | $D_2Tr_8Te_{16}P_{28}H$ | $D_2Tr_8Te_{16}P_{29}H$ | $D_2Tr_8Te_{17}P_{31}H$ | $D_2Tr_9Te_{20}P_{37}H$ | $D_2Tr_{13}Te_{22}P_{38}H$ | $D_2Tr_{13}Te_{22}P_{39}H$ | $D_4Tr_{13}Te_{22}P_{40}H$ | $D_4Tr_{13}Te_{23}P_{42}H$ | $D_6Tr_{14}Te_{27}P_{43}H$ |
|---|---|---|---|---|---|---|---|---|---|
| 1 | 1.745 | 1.801 | 1.776 | 1.783 | 1.902 | 1.916 | 1.779 | 1.985 | 1.886 |
| 2 | 1.948 | 1.788 | 1.857 | 1.990 | 1.735 | 1.742 | 1.832 | 1.974 | 1.866 |
| 3 | 1.771 | 1.889 | 1.751 | 1.824 | 1.841 | 1.869 | 1.792 | 1.803 | 1.895 |
| 4 | 1.891 | 1.776 | 1.915 | 1.822 | 1.753 | 1.747 | 1.816 | 1.881 | 1.738 |
| 5 | 1.714 | 1.753 | 1.797 | 1.980 | 1.765 | 1.779 | 1.790 | 1.814 | 1.785 |
| 6 | 1.937 | 1.818 | 1.966 | 1.959 | 1.852 | 1.879 | 1.940 | 1.868 | 1.728 |
| 7 | 1.788 | 1.932 | 1.733 | 1.768 | 1.775 | 1.785 | 1.804 | 1.812 | 1.801 |
| 8 | 1.790 | 1.792 | 1.820 | 1.800 | 1.862 | 1.922 | 1.842 | 1.960 | 1.752 |
| 9 | 1.962 | 1.785 | 1.794 | 1.968 | 1.764 | 1.714 | 1.829 | 1.809 | 1.811 |
| 10 | 1.774 | 1.813 | 1.979 | 1.736 | 1.947 | 1.958 | 1.843 | 1.937 | 1.680 |
| 11 | 1.907 | 1.823 | 1.953 | 1.821 | 1.862 | 1.980 | 1.959 | 1.777 | 1.806 |
| 12 | 1.957 | 1.960 | 1.783 | 1.797 | 1.765 | 1.768 | 1.814 | 1.860 | 1.892 |
| 13 | 1.962 | 1.957 | 1.992 | 1.789 | 1.814 | 1.784 | 1.787 | 1.792 | |
| 14 | 1.817 | 1.982 | 1.823 | 1.777 | 1.779 | 1.813 | 1.961 | 1.861 | |
| 15 | 1.793 | 1.769 | 1.730 | 1.858 | 1.787 | 1.823 | 1.772 | 1.795 | |
| 16 | 1.945 | 1.780 | 1.921 | 1.753 | 1.891 | 1.959 | 1.904 | 1.796 | |
| 17 | 1.807 | 1.882 | 1.829 | 1.917 | 1.742 | 1.891 | 1.956 | 1.787 | |
| 18 | 1.846 | 1.787 | 1.869 | | 1.759 | 1.776 | 1.769 | 1.737 | |
| 19 | 1.833 | 1.923 | 1.767 | | 1.759 | 1.755 | 1.889 | 1.753 | |
| 20 | 1.846 | 1.713 | 1.849 | | 1.853 | 1.820 | 1.713 | 1.738 | |
| 21 | 1.780 | 1.919 | | | 1.749 | 1.934 | 1.936 | 1.830 | |
| 22 | 1.836 | 1.744 | | | 1.806 | 1.794 | 1.788 | 1.765 | |
| 23 | 1.795 | 1.870 | | | 1.906 | 1.800 | 1.745 | | |
| 24 | 1.817 | 1.748 | | | 1.745 | 1.787 | 1.947 | | |
| 25 | | | | | 1.920 | | | | |
| 26 | | | | | 1.762 | | | | |
| 27 | | | | | 1.880 | | | | |
| 28 | | | | | 1.746 | | | | |

TABLE S-XVII. Binding energies (in eV) per H atom of 1 family which consists of 21 candidates in Dimer generation on 90° bent system. The configuration numbers correspond to those in Figure S8.

| Config | D |
|---|---|
| 1 | 1.125 |
| 2 | 1.302 |
| 3 | 1.307 |
| 4 | 1.142 |
| 5 | 1.753 |
| 6 | 1.200 |
| 7 | 1.792 |
| 8 | 1.200 |
| 9 | 1.319 |
| 10 | 1.783 |
| 11 | 1.848 |
| 12 | 0.941 |
| 13 | 1.202 |
| 14 | 1.793 |
| 15 | 1.202 |
| 16 | 1.322 |
| 17 | 1.311 |
| 18 | 1.144 |
| 19 | 1.757 |
| 20 | 1.130 |
| 21 | 1.306 |

TABLE S-XVIII. Binding energies (in eV) per H atom of 4 family which consists of 43 candidates in Trimer generation on 90° bent system. The configuration numbers correspond to those in Figure S8.

| Config | $D_2$Tr | $D_3$Tr | $D_4$Tr | $D_6$Tr |
|---|---|---|---|---|
| 1 | 1.552 | 1.875 | 1.607 | 1.499 |
| 2 | 1.795 | 1.608 | 1.874 | 1.527 |
| 3 | 1.637 | 1.598 | 1.650 | 1.789 |
| 4 | 1.735 | 1.653 | 1.825 | 1.565 |
| 5 | 1.596 | 1.776 | 1.606 | 1.639 |
| 6 | 1.621 | 1.586 | 1.647 | 1.631 |
| 7 | 1.651 | 1.524 | 1.564 | 1.631 |
| 8 | 1.825 | | 1.645 | 1.645 |
| 9 | 1.563 | | 1.672 | 1.574 |
| 10 | 1.826 | | | 1.795 |
| 11 | 1.652 | | | 1.535 |
| 12 | 1.596 | | | |
| 13 | 1.735 | | | |
| 14 | 1.637 | | | |
| 15 | 1.793 | | | |
| 16 | 1.552 | | | |

TABLE S-XIX. Binding energies (in eV) per H atom of 4 family which consists of 94 candidates in Tetramer generation on 90$^o$ bent system. The configuration numbers correspond to those in Figure S8.

| Config | $D_2Tr_8Te$ | $D_2Tr_9Te$ | $D_4Tr_{13}Te$ | $D_6Tr_{14}Te$ |
|---|---|---|---|---|
| 1 | 1.680 | 1.632 | 1.807 | 1.611 |
| 2 | 1.615 | 1.636 | 1.687 | 1.900 |
| 3 | 1.935 | 1.740 | 2.024 | 1.642 |
| 4 | 1.769 | 1.630 | 1.989 | 1.790 |
| 5 | 1.585 | 1.632 | 1.687 | 1.804 |
| 6 | 1.953 | 1.766 | 1.763 | 1.625 |
| 7 | 1.676 | 1.672 | 1.716 | 1.797 |
| 8 | 2.006 | 1.666 | 1.862 | 1.664 |
| 9 | 1.963 | 1.792 | 1.721 | 1.610 |
| 10 | 1.970 | 1.725 | 1.971 | 1.936 |
| 11 | 1.720 | 1.927 | 1.726 | 1.817 |
| 12 | 1.693 | 1.694 | 1.710 | 1.644 |
| 13 | 1.929 | 1.764 | 1.938 | 1.858 |
| 14 | 1.725 | 1.723 | 1.668 | 1.649 |
| 15 | 1.793 | 1.972 | 1.846 | 1.893 |
| 16 | 1.666 | 1.655 | 1.694 | 1.610 |
| 17 | 1.672 | 1.654 | 1.724 | |
| 18 | 1.767 | 1.959 | 1.829 | |
| 19 | 1.631 | 2.004 | 1.661 | |
| 20 | 1.737 | 1.672 | 1.694 | |
| 21 | 1.634 | 1.830 | 1.627 | |
| 22 | | 1.692 | 1.758 | |
| 23 | | 1.676 | 1.663 | |
| 24 | | 1.953 | | |
| 25 | | 1.586 | | |
| 26 | | 1.769 | | |
| 27 | | 1.677 | | |
| 28 | | 1.935 | | |
| 29 | | 1.614 | | |
| 30 | | 1.677 | | |
| 31 | | 1.624 | | |
| 32 | | 1.759 | | |
| 33 | | 1.611 | | |

TABLE S-XX. Binding energies (in eV) per H atom of 6 family which consists of 92 candidates in Pentamer generation on 90$^o$ bent system. The configuration numbers correspond to those in Figure S8.

| Config | $D_2Tr_8Te_{16}P$ | $D_2Tr_8Te_{17}P$ | $D_2Tr_9Te_{21}P$ | $D_4Tr_{13}Te_{23}P$ | $D_4Tr_{13}Te_{24}P$ | $D_6Tr_{14}Te_{28}P$ |
|---|---|---|---|---|---|---|
| 1 | 1.769 | 1.949 | 1.852 | 1.861 | 1.924 | 1.729 |
| 2 | 1.962 | 1.830 | 1.852 | 1.812 | 1.772 | 1.895 |
| 3 | 1.837 | 1.846 | 1.893 | 1.892 | 1.904 | 1.862 |
| 4 | 1.757 | 1.866 | 1.896 | 1.840 | 1.787 | 1.756 |
| 5 | 1.933 | 1.831 | 1.898 | 1.884 | 1.746 | 1.918 |
| 6 | 1.827 | 1.836 | 1.893 | 1.860 | 1.761 | 1.787 |
| 7 | 1.845 | 2.014 | 1.868 | 2.008 | 1.784 | 1.765 |
| 8 | 1.895 | 1.871 | 1.894 | 1.864 | 1.725 | 1.794 |
| 9 | 1.997 | 1.897 | 1.835 | 1.843 | 1.766 | 1.792 |
| 10 | 2.010 | 1.896 | 2.013 | 1.895 | 1.695 | 1.925 |
| 11 | 1.867 | 1.902 | 1.845 | 1.995 | 1.766 | 1.763 |
| 12 | 1.895 | 1.818 | 1.867 | 1.837 | 1.796 | 1.866 |
| 13 | 1.838 | 1.854 | 1.831 | 1.756 | | 1.902 |
| 14 | 1.881 | 1.815 | 1.949 | 1.934 | | 1.734 |
| 15 | 1.858 | | 1.829 | 1.825 | | |
| 16 | 1.860 | | 1.783 | 1.776 | | |
| 17 | 1.811 | | | 1.790 | | |
| 18 | | | | 1.770 | | |
| 19 | | | | 1.964 | | |

TABLE S-XXI. Binding energies (in eV) per H atom of 8 family which consists of 167 candidates in Hexamer generation on 90$^o$ bent system. The configuration numbers correspond to those in Figure S8.

| Config | $D_2Tr_8Te_{16}P_{29}H$ | $D_2Tr_8Te_{16}P_{30}H$ | $D_2Tr_8Te_{17}P_{31}H$ | $D_2Tr_9Te_{21}P_{32}H$ | $D_4Tr_{13}Te_{23}P_{33}H$ | $D_4Tr_{13}Te_{23}P_{34}H$ | $D_4Tr_{13}Te_{24}P_{36}H$ | $D_6Tr_{14}Te_{28}P_{37}H$ |
|---|---|---|---|---|---|---|---|---|
| 1 | 1.836 | 1.895 | 1.869 | 1.894 | 2.018 | 1.868 | 2.101 | 1.937 |
| 2 | 2.045 | 1.874 | 1.958 | 2.103 | 1.833 | 1.930 | 2.079 | 1.912 |
| 3 | 1.863 | 1.992 | 1.843 | 1.926 | 1.965 | 1.881 | 1.903 | 1.961 |
| 4 | 1.990 | 1.869 | 2.021 | 1.935 | 1.836 | 1.912 | 1.989 | 1.793 |
| 5 | 1.809 | 1.855 | 1.905 | 2.091 | 1.875 | 1.894 | 1.915 | 1.837 |
| 6 | 2.035 | 1.915 | 2.074 | 2.080 | 1.989 | 2.045 | 1.982 | 1.778 |
| 7 | 1.884 | 2.034 | 1.836 | 1.873 | 1.882 | 1.907 | 1.916 | 1.867 |
| 8 | 1.895 | 1.895 | 1.927 | 1.908 | 2.024 | 1.951 | 2.065 | 1.810 |
| 9 | 2.069 | 1.883 | 1.895 | 2.076 | 1.808 | 1.926 | 1.913 | 1.872 |
| 10 | 1.879 | 1.922 | 2.090 | 1.838 | 2.070 | 1.948 | 1.958 | 1.773 |
| 11 | 2.020 | 1.926 | 2.078 | 1.929 | 2.089 | 2.066 | 1.870 | 1.864 |
| 12 | 2.072 | 2.066 | 1.894 | 1.896 | 1.876 | 1.916 | 1.962 | 1.945 |
| 13 | 2.068 | 2.071 | 2.104 | 1.899 | 1.882 | 1.894 | 1.886 | |
| 14 | 1.916 | 2.090 | 1.926 | 1.871 | 1.922 | 2.065 | 1.962 | |
| 15 | 1.899 | 1.878 | 1.828 | 1.957 | 1.926 | 1.878 | 1.886 | |
| 16 | 2.049 | 1.876 | 2.027 | 1.845 | 2.064 | 2.018 | 1.900 | |
| 17 | 1.910 | 1.990 | 1.927 | 2.022 | 1.992 | 2.071 | 1.889 | |
| 18 | 1.953 | 1.883 | 1.982 | | 1.869 | 1.860 | 1.838 | |
| 19 | 1.930 | 2.025 | 1.863 | | 1.856 | 1.989 | 1.854 | |
| 20 | 1.950 | 1.832 | 1.948 | | 1.917 | 1.809 | 1.842 | |
| 21 | 1.871 | 2.020 | | | 2.036 | 2.034 | 1.931 | |
| 22 | 1.933 | 1.835 | | | 1.896 | 1.883 | 1.865 | |
| 23 | 1.884 | 1.967 | | | 1.896 | 1.835 | | |
| 24 | 1.915 | 1.839 | | | 1.874 | 2.043 | | |

## III. DFT-DFTB COMPARISON

In DFT calculations, we use SIESTA package[52] with LDA as the exchange-correlation functional[53]. Double-$\zeta$ polarized orbitals are used as the basis set. Mesh cutoff value of 500 Ry is used. The convergence criterion for optimization of atomic positions is set to be $2{\times}10^{-3}$ eV/Ang.

### A. Dimer Configurations

Ten dimer configurations are tested by adding a second hydrogen atom up to the $5^{th}$ nearest neighbour of the initial site.[32]. All-tested configurations are depicted in Figure S9 and total relative energies with respect to the minimum energy configuration are given in Table S-XXII for three cases. In the first two cases all hydrogen and carbon atoms are relaxed. In the third case, the hydrogen atom and the four nearest carbon atoms are relaxed while the remaining carbons are fixed. This constraint is meant to mimic the effect of an underlying substrate. DFT calculations predict ortho-positioned Conf-1 and para-positioned Conf-3 as the most favourable configurations, consistent with Ref. 32, while DFTB calculations yield a slight difference in energies. Also DFT-constrained calculations give the same ordering for the first two stable configurations as DFT-relaxed calculations. Comparison of DFT and DFTB results is given in Figure S14(a).

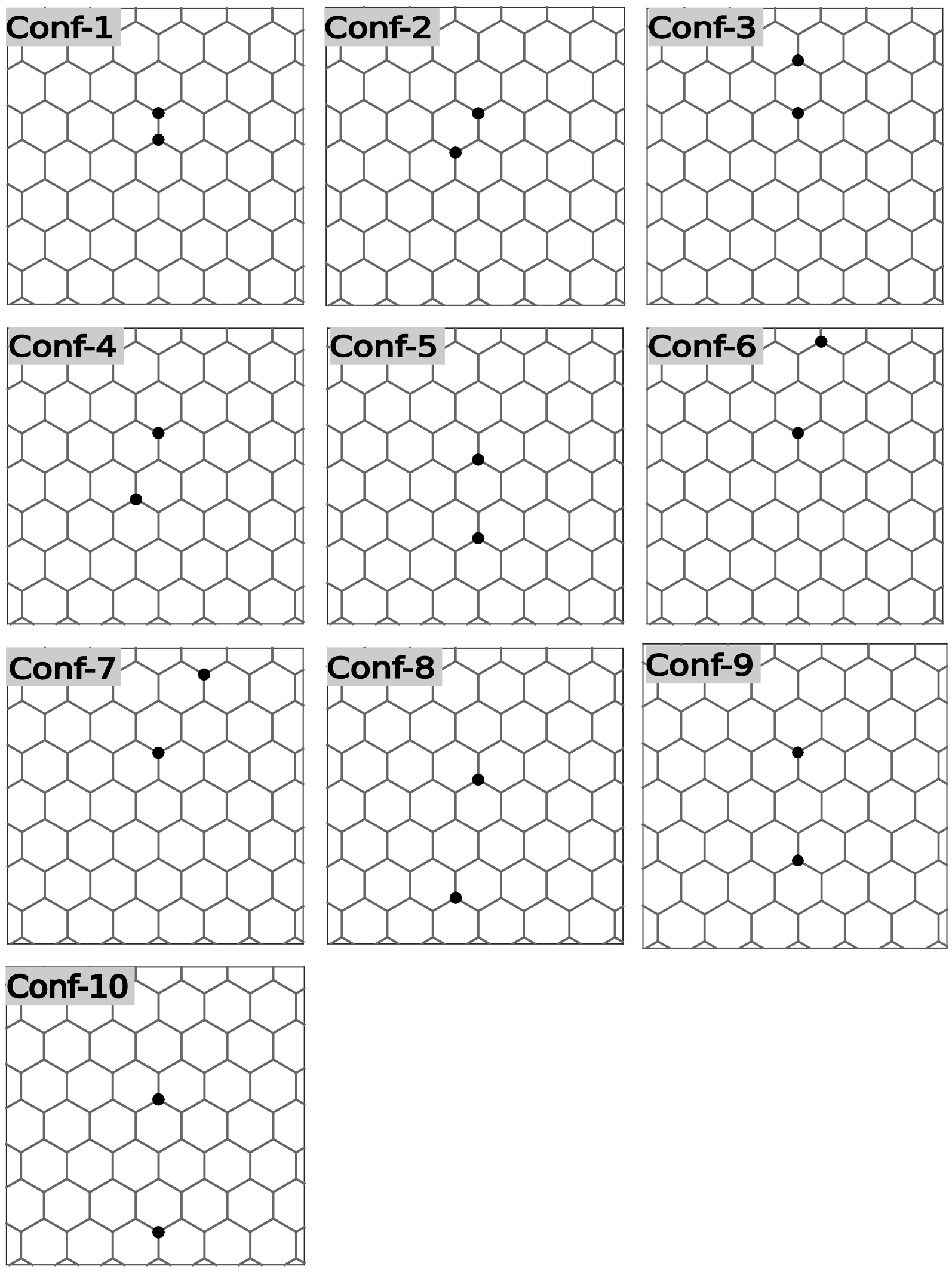

FIG. S9. Dimer configurations which are determined for checking DFT-DFTB consistency on flat graphene.

TABLE S-XXII. Relative total energies (in meV) with respect to the configuration with minimum energy based on DFT-relaxed, DFTB-relaxed and DFT-constrained results for dimer configurations.

Relative Total Energies of Dimer Family (7×7×1 Super Cell)

| Config | DFT | DFTB | DFT-constrained |
|---|---|---|---|
| Conf-1 | 18.740 | 0.000 | 0.000 |
| Conf-2 | 1259.657 | 1270.100 | 1255.175 |
| Conf-3 | 0.000 | 63.400 | 11.662 |
| Conf-4 | 873.090 | 959.400 | 890.611 |
| Conf-5 | 1267.330 | 1310.500 | 1206.037 |
| Conf-6 | 552.061 | 645.3 | 595.783 |
| Conf-7 | 1175.140 | 1231.200 | 1134.140 |
| Conf-8 | 706.789 | 758.300 | 712.862 |
| Conf-9 | 1179.084 | 1219.300 | 1139.260 |
| Conf-10 | 598.299 | 718.000 | 634.158 |

### B. Trimer Configurations

9 configurations are tested by addition of third hydrogen atom to dimers for three cases as presented in Figure S10. According to the DFT results (see TableS-XXIII), Conf-4, Conf-1 and Conf-7 are the most favourable configurations, in agreement with Ref. 31. DFTB and DFT-constrained calculations follow the same ordering in total energies. Comparison of DFT and DFTB results is given in Figure S14(b) for trimers.

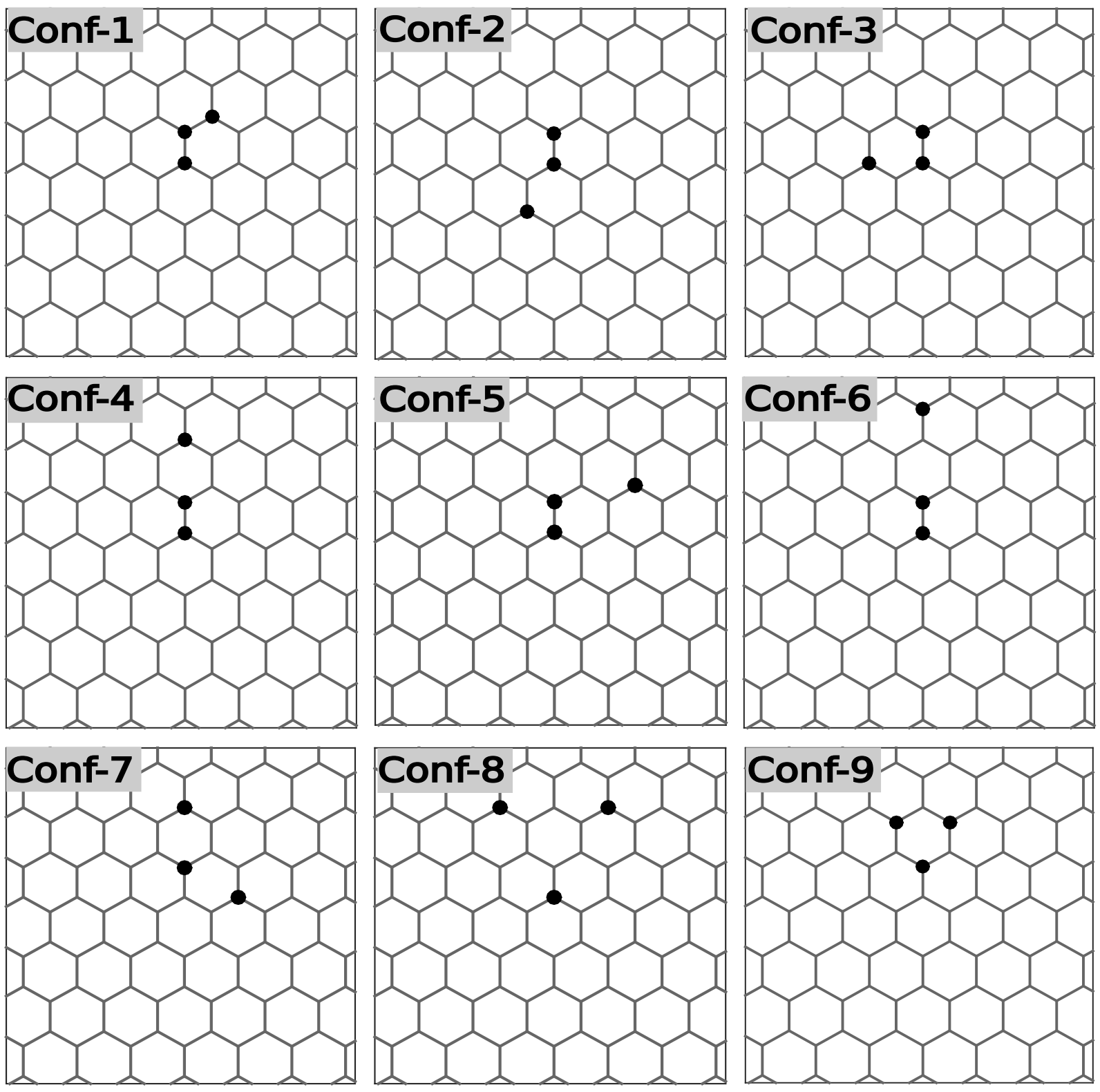

FIG. S10. Trimer configurations which are determined for checking DFT-DFTB consistency on flat graphene.

TABLE S-XXIII. Relative total energies (in meV) with respect to the minimum energy configuration, based on DFT-relaxed, DFTB-relaxed and DFT-constrained results for trimer configurations.

Relative Total Energies of Trimer Family (7×7×1 Super Cell)

| Config | DFT | DFTB | DFT constrained |
|---|---|---|---|
| Conf-1 | 101.038 | 102.100 | 56.004 |
| Conf-2 | 689.914 | 667.500 | 781.943 |
| Conf-3 | 490.893 | 506.300 | 0.509950 |
| Conf-4 | 0.000 | 0.000 | 0.000 |
| Conf-5 | 596.696 | 643.200 | 612.919 |
| Conf-6 | 726.108 | 769.700 | 717.822 |
| Conf-7 | 101.068 | 149.000 | 145.940 |
| Conf-8 | 1836.565 | 1974.10 | 1803.334 |
| Conf-9 | 2114.458 | 2103.40 | 2191.332 |

### C. Tetramer Configurations

12 configurations are simulated as illustrated in Figure S11. Conf-3, Conf-4, Conf-6 and Conf-5 are the most stable configurations according to relaxed DFT calculations, respectively (see Table S-XXIV). DFT, DFTB and DFT-constrained results are compatible with each other and also in agreement with the literature except for the order of the first four favourable configurations. Conf-8 can be included in the list of stable configurations because of the slight energy difference with the threshold value. Conf-2, Conf-7 and Conf-8 are included in the favourable configurations because their relative total energies are quite close to threshold value of 200 meV.

TABLE S-XXIV. Relative total energies (in meV) with respect to the minimum energy configuration based on DFT-relaxed, DFTB-relaxed and DFT-constrained simulations for tetramer configurations.

Relative Total Energies of Tetramer Family (7×7×1 Super Cell)

| Config | DFT | DFTB | DFT constrained |
|---|---|---|---|
| Conf-1 | 1008.733 | 1064.700 | 934.218 |
| Conf-2 | 254.052 | 278.700 | 270.930 |
| Conf-3 | 0.000 | 0.000 | 0.000 |
| Conf-4 | 62.131 | 121.100 | 133.706 |
| Conf-5 | 120.355 | 148.600 | 166.771 |
| Conf-6 | 86.011 | 66.900 | 104.084 |
| Conf-7 | 221.922 | 250.700 | 294.772 |
| Conf-8 | 219.266 | 303.300 | 327.679 |
| Conf-9 | 632.249 | 591.000 | 788.747 |
| Conf-10 | 719.338 | 773.200 | 803.571 |
| Conf-11 | 760.456 | 867.500 | 957.656 |
| Conf-12 | 936.271 | 971.700 | 996.032 |

### D. Pentamer Configurations

As shown in Figure S12, 16 configuration are investigated for Pentamers. Conf-5, Conf-3, Conf-1, Conf-2 and Conf-4 are the most favourable configurations based on DFT-relaxed calculations. (see Table S-XXV). Order of Conf-3 and Conf-1 changes in DFTB calculations with approximately 22 meV of relative energy difference, which does not affect our stability considerations.

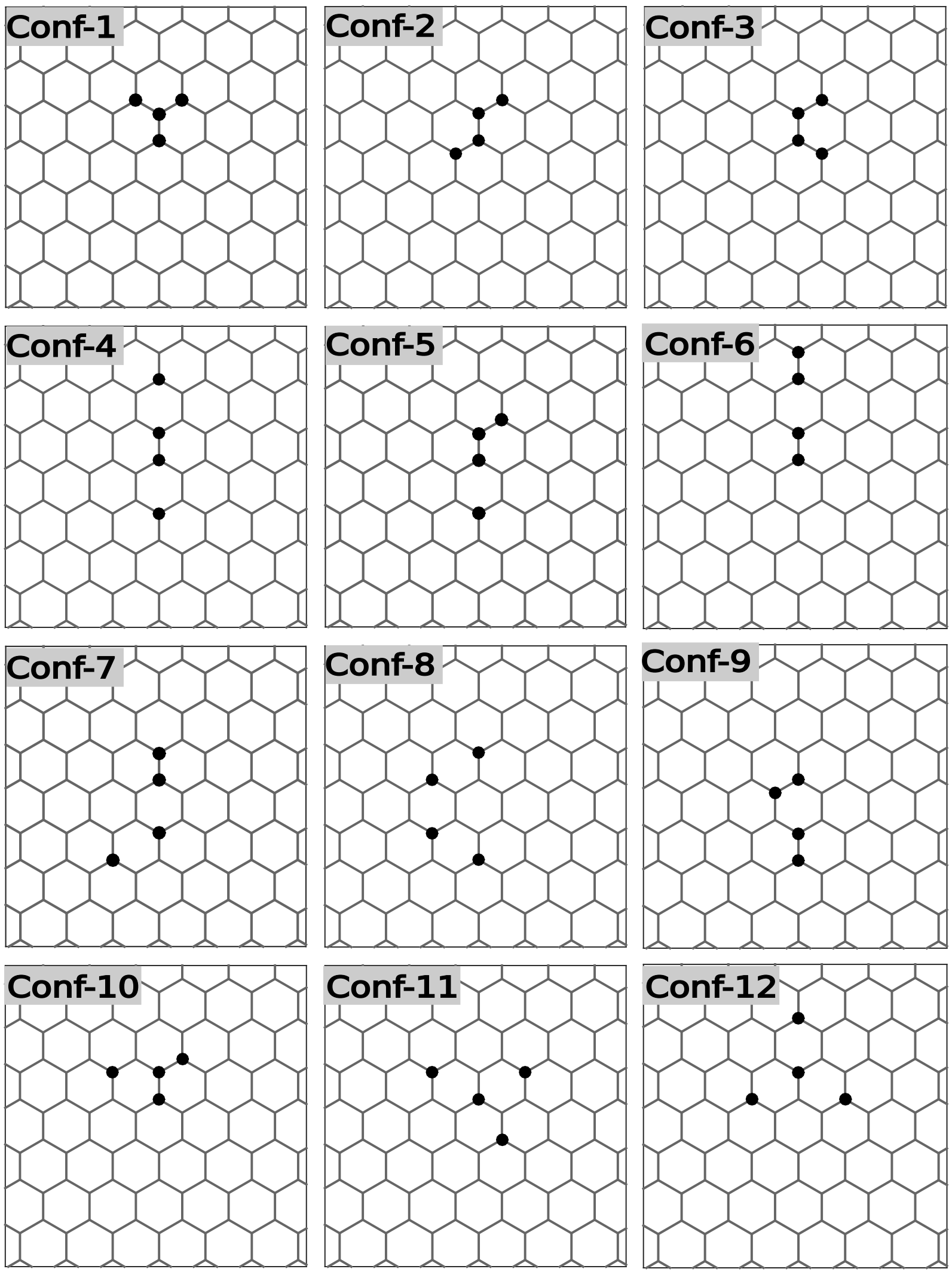

FIG. S11. Tetramer configurations used determined for checking DFT-DFTB consistency on flat graphene.

### E. Hexamer Configurations

13 configurations are studied as depicted in Figure S13. Conf-2 is found to be the most preferable configuration from DFT-relaxed, DFTB and DFT-constrained simulations, as the configuration has a special hexagonal symmetry. DFT results reveal that Conf-2 and Conf-12 are the most preferable configurations. In Ref. 31 Conf-2 is the second favourable configuration whereas Conf-12 is not included in the study. The results of DFTB calculations are even more interesting. Conf-2 and Conf-12 are the most stable configurations and Conf-1 and Conf-13 are also found to be stable. Conf-1 was qualified as the most preferable configuration in Ref 31 but Conf-13, which also indicates single line formation with unpaired hydrogen atoms on flat graphene, was not included. Relative total energies of all configurations belonging to hexamer family can be found in TableS-XXVI. We note that, relative total energies of DFT calculations are higher than those of DFTB calculations in general according to the Figure S14(e). These results confirm that formation of hydrogen lines on flat and curved graphene should be investigated in more detail.

### F. Adsorption on Curved Graphene

In Figure S15, total energies of single hydrogen adsorbed curved graphene are plotted for different adsorption sites. Bothe the minimum energy adsorption site and the range of energies are in agreement with DFTB simulations (see

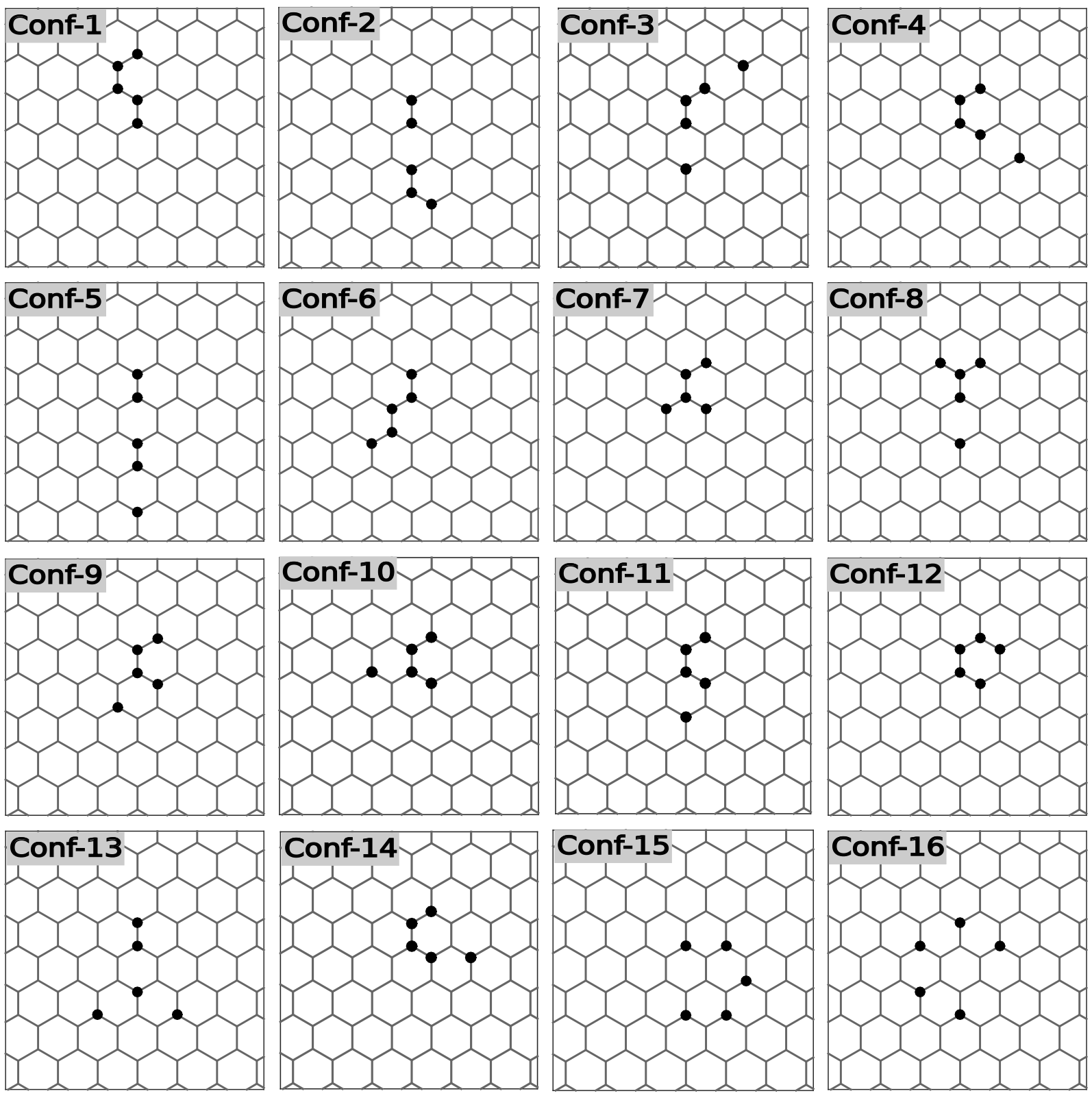

FIG. S12. Pentamer configurations which are used for checking DFT-DFTB consistency on flat graphene.

TABLE S-XXV. Relative energies (in meV) with respect to the minimum energy configuration based on DFT-relaxed, DFTB-relaxed and DFT-constrained results for pentamer configurations.

Hydrogen Pentamer Energies (7×7×1 Super Cell)

| Config | DFT | DFTB | DFT constrained |
|---|---|---|---|
| Conf-1 | 46.242 | 57.500 | 0.000 |
| Conf-2 | 73.668 | 79.000 | 66.335 |
| Conf-3 | 31.082 | 64.30 | 94.882 |
| Conf-4 | 105.009 | 97.200 | 91.463 |
| Conf-5 | 0.000 | 0.000 | 32.946 |
| Conf-6 | 249.879 | 301.400 | 243.222 |
| Conf-7 | 265.509 | 332.500 | 216.618 |
| Conf-8 | 333.301 | 415.300 | 340.389 |
| Conf-9 | 425.241 | 450.10 | 503.640 |
| Conf-10 | 440.795 | 459.200 | 496.510 |
| Conf-11 | 447.319 | 461.900 | 473.437 |
| Conf-12 | 680.264 | 621.000 | 665.817 |
| Conf-13 | 416.972 | 437.300 | 481.684 |
| Conf-14 | 549.854 | 598.600 | 617.269 |
| Conf-15 | 4003.330 | 4311.300 | 3051.663 |
| Conf-16 | 293.239 | 416.500 | 387.458 |

Figure 5).

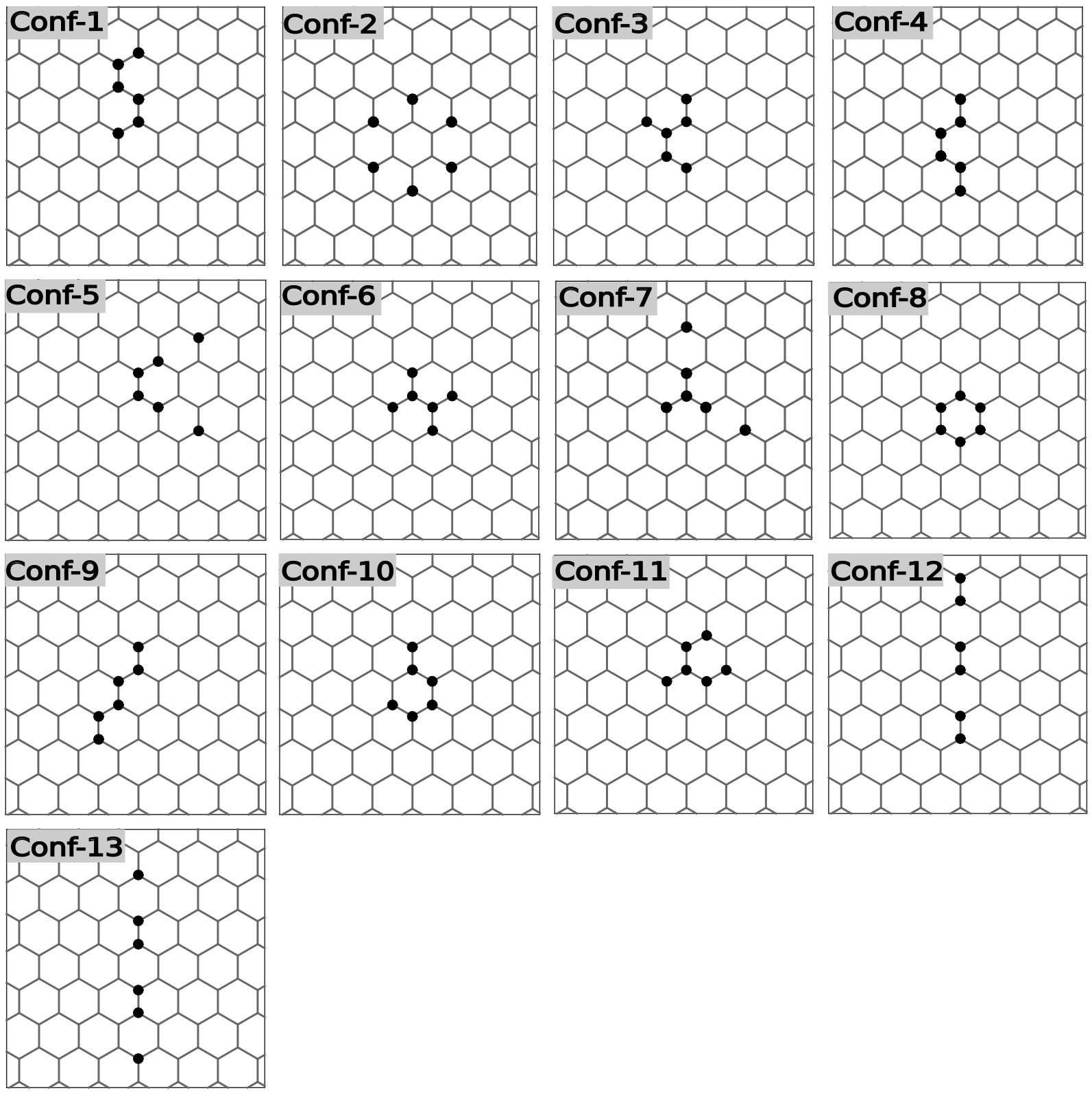

FIG. S13. Hexamer configurations which are used for checking DFT-DFTB consistency on flat graphene.

TABLE S-XXVI. Relative energies (in meV) with respect to the minimum energy configuration based on DFT-relaxed, DFTB-relaxed and DFT-constrained results for hexamer configurations.

Hydrogen Hexamer Energies (7×7×1 Super Cell)

| Config | DFT | DFTB | DFT constrained |
|---|---|---|---|
| Conf-1 | 215.542 | 132.8 | 108.271 |
| Conf-2 | 0.000 | 0.000 | 0.000 |
| Conf-3 | 293.734 | 248.4 | 116.738 |
| Conf-4 | 323.936 | 228.8 | 161.667 |
| Conf-5 | 324.979 | 215.8 | 206.717 |
| Conf-6 | 550.778 | 571.3 | 361.337 |
| Conf-7 | 446.704 | 413.4 | 402.267 |
| Conf-8 | 743.867 | 475.8 | 673.822 |
| Conf-9 | 592.909 | 542.7 | 505.188 |
| Conf-10 | 1001.771 | 831.8 | 933.741 |
| Conf-11 | 1243.242 | 1129.8 | 1138.191 |
| Conf-12 | 156.209 | 5.2 | 1363.930 |
| Conf-13 | 262.181 | 186.2 | 214.896 |

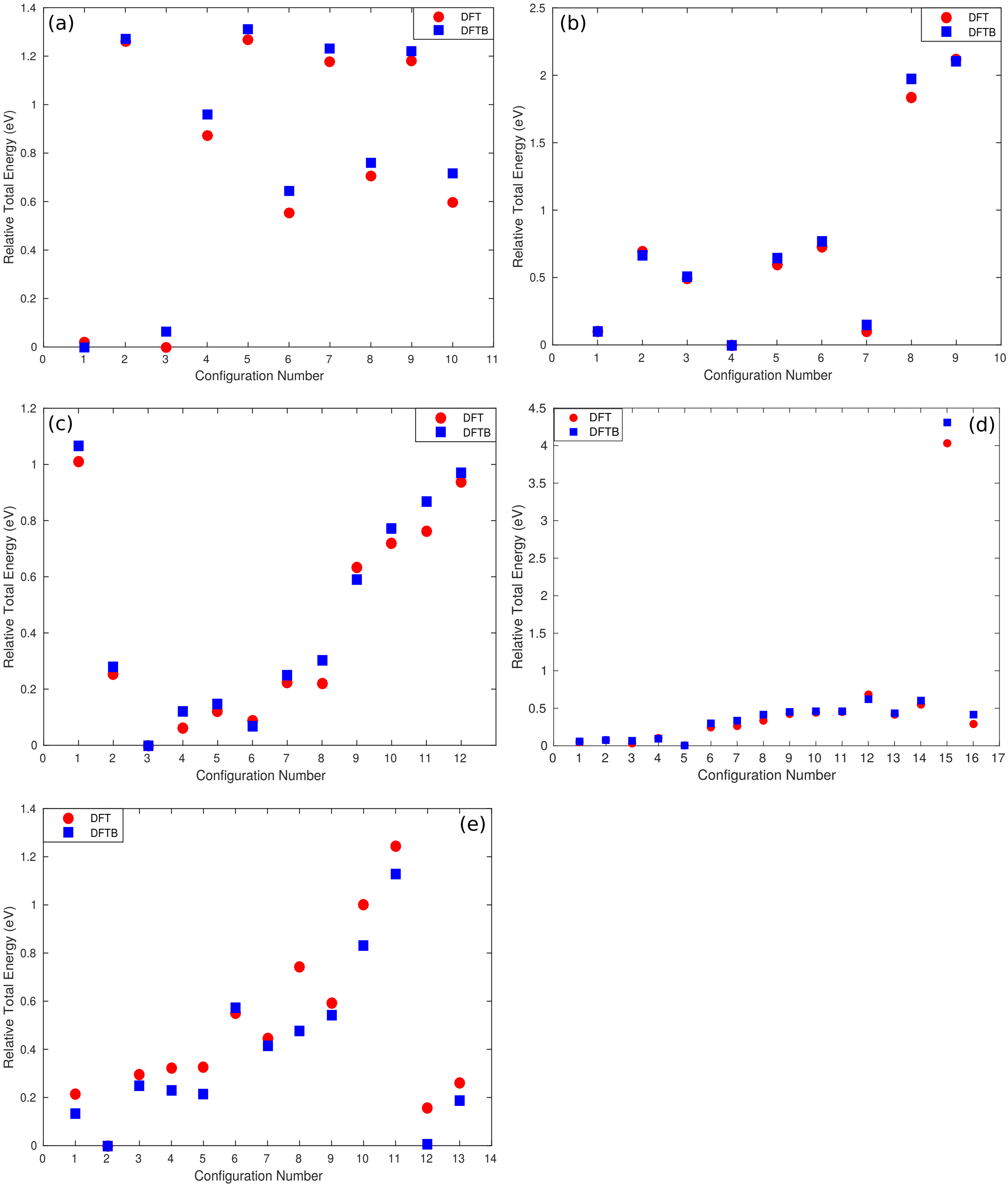

FIG. S14. DFT-DFTB comparison on selected configurations from the literature. These selected configurations are illustrated in Figs.S9..S13

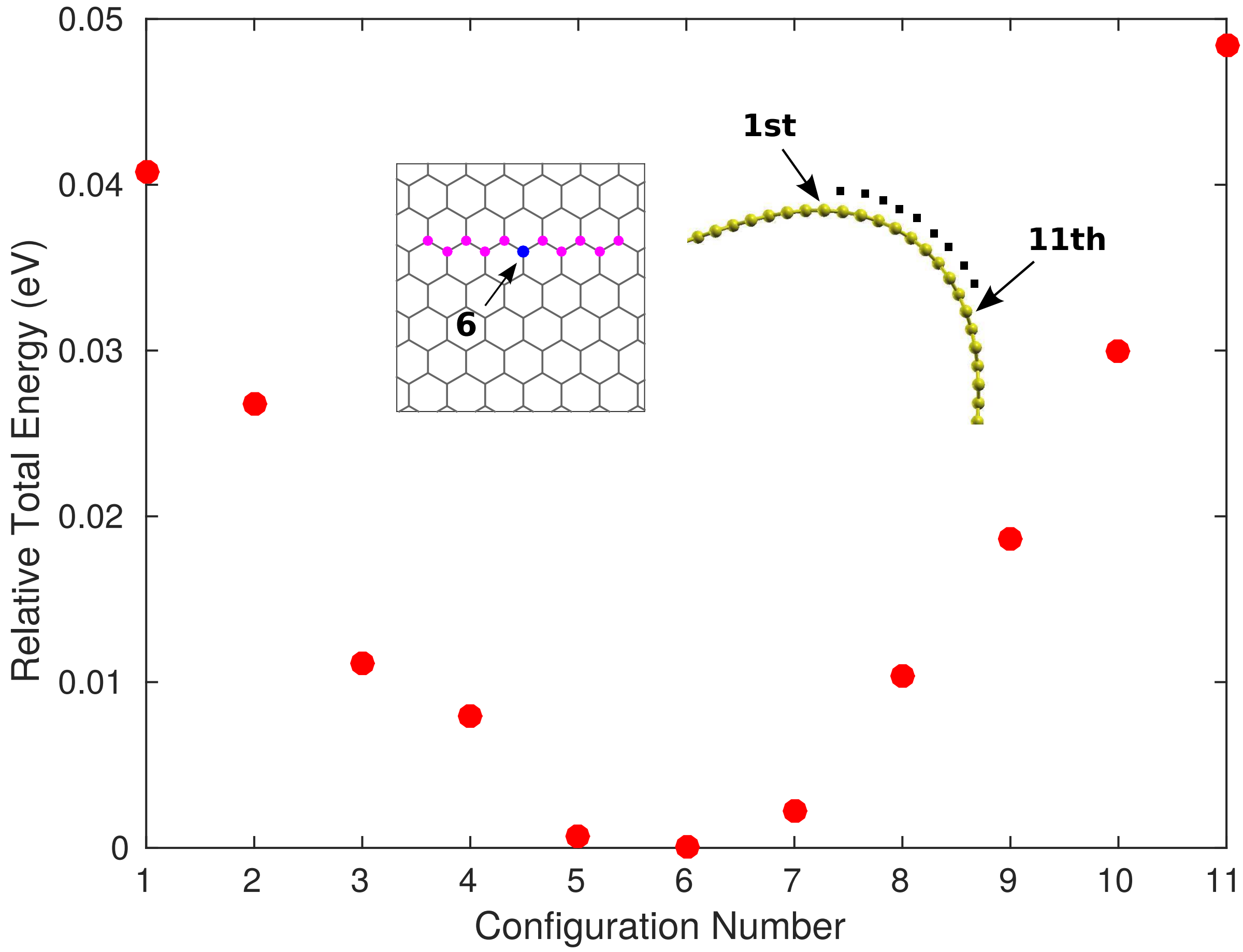

FIG. S15. Energetics of single hydrogen adsorption on curved graphene with 90°-bending based on SIESTA calculations.